\documentclass[12pt, a4paper]{article}
\baselineskip=\normalbaselineskip

\usepackage{mathrsfs,amsbsy,amssymb,latexsym,amsfonts,amsmath,amsthm}
\usepackage{graphicx,color}
\usepackage[nosort]{cite}
\usepackage[export]{adjustbox}

\usepackage{bm}

\newcommand{\bbZ}{{\mathbb{Z}}}

\makeatletter
\catcode`\@=11
\@addtoreset{equation}{section}

\def\@seccntformat#1{\csname the#1\endcsname.~~}
\makeatother

\begin{document}

\begin{titlepage} 

\renewcommand{\thefootnote}{\fnsymbol{footnote}}
\begin{flushright}
%   KUNS-2887, RBRC-****
  KUNS-2887
\end{flushright}
\vspace*{1.0cm}

\begin{center}
{\Large \bf
Tensor network approach to 2D Yang-Mills theories
}
\vspace{1.0cm}

\centerline{
{Masafumi Fukuma${}^1$}%
% {Masafumi Fukuma}%
\footnote{
  E-mail address: fukuma@gauge.scphys.kyoto-u.ac.jp
}, 
{Daisuke Kadoh${}^{2,3}$}%
% {Masafumi Fukuma}%
\footnote{
  E-mail address: dkadoh@mail.doshisha.ac.jp, kadoh@keio.jp
}
and
{Nobuyuki Matsumoto${}^4$}%
% {Nobuyuki Matsumoto}%
\footnote{
  E-mail address: nobuyuki.matsumoto@riken.jp} 
}

\vskip 0.8cm
  ${}^1${\it Department of Physics, Kyoto University,
  Kyoto 606-8502, Japan}
\vskip 0.1cm
  ${}^2${\it Faculty of Sciences and Engineering, 
  Doshisha University, %Kyotanabe, 
  Kyoto 610-0394, Japan}
\vskip 0.1cm
  ${}^3${\it Research and Educational Center for Natural Sciences, \\
  Keio University, Yokohama 223-8521, Japan}
\vskip 0.1cm
  ${}^4${\it RIKEN/BNL Research center, Brookhaven National Laboratory,
  Upton, NY 11973, USA}
\vskip 1.2cm

\end{center}

%%%%%%%%%%%%%%%%%%%%%%%%%%%%%%%%%%%%%%%
\begin{abstract}
%%%%%%%%%%%%%%%%%%%%%%%%%%%%%%%%%%%%%%%

We propose a novel tensor network representation 
for two-dimensional Yang-Mills theories 
with arbitrary compact gauge groups. 
In this method, 
tensor indices are directly given by group elements 
with no direct use of the character expansion. 
We apply the tensor renormalization group method 
to this tensor network for $SU(2)$ and $SU(3)$, 
and find that the free energy density and the energy density 
are accurately evaluated.
We also show that the singular value decomposition of a tensor 
has a group theoretic structure 
and can be associated with the character expansion. 

%%%%%%%%%%%%%%%%%%%%%%%%%%%%%%%%%%%%%%% 
\end{abstract}
%%%%%%%%%%%%%%%%%%%%%%%%%%%%%%%%%%%%%%% 
\end{titlepage}

\pagestyle{empty}
\pagestyle{plain}

\tableofcontents
\setcounter{footnote}{0}

%%%%%%%%%%%%%%%%%%%%%%%%%%%%%%%%%%%%%%%
%%%%%%%%%%%%%%%%%%%%%%%%%%%%%%%%%%%%%%%
\section{Introduction}
\label{sec:Introduction}
%%%%%%%%%%%%%%%%%%%%%%%%%%%%%%%%%%%%%%%
%%%%%%%%%%%%%%%%%%%%%%%%%%%%%%%%%%%%%%%

The tensor network (TN) method
\cite{Niggemann:1997cq, Verstraete:2004cf,Levin:2006jai,Xie:2009zzd,Gu:2010yh} 
is an attractive approach for studying many body systems, 
because it is free from the sign problem in the first place,%
\footnote{%-----
  Recently, significant progress has been made 
  also in the Monte Carlo (MC) approach to the sign problem
  \cite{Parisi:1983cs,Klauder:1983sp,Aarts:2009dg,Nishimura:2015pba,
  Witten:2010cx,Cristoforetti:2012su,Cristoforetti:2013wha,
  Fujii:2013sra,Alexandru:2015sua,Fukuma:2017fjq,Alexandru:2017oyw,
  Fukuma:2019wbv,Fukuma:2019uot,Fukuma:2020fez,Fukuma:2021aoo,
  Mori:2017pne,Mori:2017nwj,Alexandru:2018fqp}, 
  giving rise to a hope 
  that MC simulations can be performed at a reasonable computational cost. 
  Two approaches (TN and MC) may play complementary roles in the future.
} %-------------
and has a potential to precisely investigate critical phenomena 
in the large volume limit. 
In field theory, 
the tensor renormalization group (TRG) method 
\cite{Levin:2006jai}  
and its variations \cite{Xie:2012zzz, Adachi:2019paf, Kadoh:2019kqk}
are widely used to study various models 
such as the Schwinger model
\cite{Shimizu:2014uva,Shimizu:2014fsa,Shimizu:2017onf,Butt:2019uul}, 
the Gross-Neveu and NJL models \cite{Takeda:2014vwa, Akiyama:2020soe},
scalar field theories 
\cite{Kadoh:2018tis, Kadoh:2019ube, Akiyama:2020ntf,Akiyama:2021zhf},
the Yang-Mills and gauge-Higgs models \cite{Asaduzzaman:2019mtx, Bazavov:2019qih}, 
the Wess-Zumino model \cite{Kadoh:2018hqq}, 
and other related models \cite{Kawauchi:2016xng,Akiyama:2020sfo,Kadoh:2021fri}. 

For gauge groups $U(1)$ 
\cite{Shimizu:2014uva,Shimizu:2014fsa,Shimizu:2017onf,
Kawauchi:2016xng,Kuramashi:2019cgs}
and $SU(2)$ 
\cite{Asaduzzaman:2019mtx,Bazavov:2019qih}, 
the character expansion was employed 
to represent the partition function with a tensor network. 
However, since the character expansion becomes a demanding task 
for higher-rank gauge groups, 
it remains as a difficult issue to apply the TN method 
to $SU(N)$ gauge theory for $N \ge 3$ including QCD.

In this paper, we propose a novel method 
to create a tensor network for two-dimensional Yang-Mills theory 
with no direct use of the character expansion. 
The Haar measure is discretized, 
and the group integration is replaced 
by a summation over $K$ randomly generated configurations. 
Then, the plaquette is regarded as a rank-$4$ tensor 
whose index runs from $1$ to $K$, 
and the set of plaquettes constitutes a tensor network.  
We test our method for $SU(2)$ and $SU(3)$ gauge groups, 
and find that the free energy density and the energy density 
agree very well with exact results. 
We also clarify the mathematical structure behind our method. 

This paper is organized as follows.
In Sec.~\ref{sec:2dYM_TRG}, 
we introduce our tensor network representation 
for two-dimensional Yang-Mills theories 
with arbitrary compact gauge groups $G$, 
and discuss its relation with the character expansion. 
In Sec.~\ref{sec:numerical_results}, 
we test our method for $G=SU(2)$ and $G=SU(3)$. 
Section~\ref{sec:discussion} is devoted to summary and discussion. 
Appendices provide some useful formulas in group theory.

%%%%%%%%%%%%%%%%%%%%%%%%%%%%%%%%%%%%%%%
%%%%%%%%%%%%%%%%%%%%%%%%%%%%%%%%%%%%%%%
\section{Tensor network representations for 2D Yang-Mills theories}
\label{sec:2dYM_TRG}
%%%%%%%%%%%%%%%%%%%%%%%%%%%%%%%%%%%%%%%
%%%%%%%%%%%%%%%%%%%%%%%%%%%%%%%%%%%%%%%

In this section, 
we introduce a new tensor network representation 
for two-dimensional Yang-Mills theories, 
and discuss its relation with the character expansion. 
We exclusively consider pure Yang-Mills theory for simplicity.
It is straightforward to extend our method 
to systems with interacting matter fields. 

%%%%%%%%%%%%%%%%%%%%%%%%%%%%%%%%%%%%%%%
\subsection{Method}
\label{sec:method}
%%%%%%%%%%%%%%%%%%%%%%%%%%%%%%%%%%%%%%%

We consider the Yang-Mills theory with a compact gauge group $G$
on an infinite lattice 
$\Gamma \equiv \{n=(n_1,n_2)\,|\, n_\mu \in \bbZ~(\mu=1,2) \}$. 
The lattice spacing $a$ is set to $a=1$ unless otherwise noted, 
and $\hat \mu$ is the unit vector in the $\mu$ direction. 

Let $U_\mu(n)$ be the $G$-valued link field 
on links $(n,n+\hat \mu)$.
The lattice action (the Wilson action) is given by% 
\footnote{ %-----
  We obtain the usual continuum action 
  for $\beta=2N/(ga)^2$ 
  with $U_P (n) \simeq {\rm exp}(i a^2 F_{12}(n))$
  in the naive continuum limit.
} %--------------
\begin{align}
  S= \frac{\beta}{N}\,\sum_{n \in \Gamma} {\rm Re \, tr}\,[1-U_P(n)], 
\end{align}
where $U_P(n)$ is the plaquette field, 
\begin{align}
  U_P (n) = U_{1}(n)\, U_2(n+\hat 1)\, 
  U_1^\dag(n+\hat 2)\, U_2^\dag(n).  
\end{align}
The partition function is defined as
$Z= \int DU\, e^{-S} $,
where
$DU \equiv \prod_{n \in \Gamma}  dU_1(n)\, dU_2(n)$
with $dU$ the Haar measure of $G$.
Note that the partition function $Z$ can be written 
in the form of a tensor network 
with indices continuously taking values in $G$,
\begin{align}
  Z = {\mathfrak T}{\mathfrak r} \prod_{n\in\Gamma} 
  {\mathfrak T}_{g(n) h(n) g'(n) h'(n)}, 
\label{inf_Z}
\end{align}
where 
\begin{align}
  {\mathfrak T}_{g_1 g_2 g_3 g_4} 
  = e^{-(\beta/N)\, {\rm Re\,tr\,}(1 - g_1 g_2 g_3^{\dag} g_4^{\dag} )}
\label{inf_tensor}
\end{align}
and ${\mathfrak T}{\mathfrak r}$ stands for the group integrations 
for $g(n),h(n)\in G$ $(n \in \Gamma)$ 
under a proper identification of indices.%
\footnote{ %-----
  We make the identifications $g'(n)=g(n+\hat{2})$ and $h'(n)=h(n-\hat{1})$. 
} %-----

\begin{figure}[htbp]
  \centering
  \includegraphics[width=110mm]{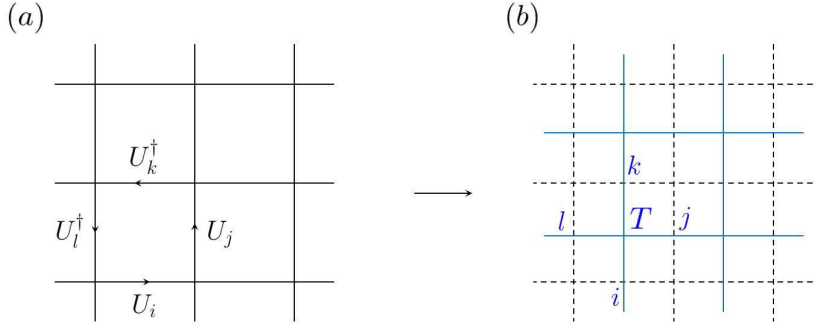}
  \caption{
  Two-dimensional square lattice. 
  (a) A plaquette variable consisting of $U_i,U_j,U_k,U_l$. 
  (b) The corresponding tensor \eqref{tensor} 
  assigned to the center of the plaquette. 
}
\label{fig:tensor}
\end{figure}

We now discretize the Haar measure $dU$ 
to represent $Z$ as a tensor network 
with indices in a finite range: 
\begin{align}
  \int dU\, f(U) \approx \frac{1}{K}\,\sum_{i=1}^K f(U_i), 
\label{approximation}
\end{align}
where $\mathring {G}=\{U_1,U_2,\ldots,U_K\}$ 
consists of random points 
uniformly chosen from the group manifold.
Applying Eq.~\eqref{approximation} to the Haar measures in $DU$ 
leads to
\begin{align}
  Z \approx {\rm Tr} \prod_{n \in \Gamma}  T_{i_nj_ni_n'j_n'},
\label{TN}
\end{align}
where
\begin{align}
  T_{ijkl} = \frac{1}{K^2} e^{-(\beta/N)\,
  {\rm Re\,tr\,}(1 - U_i U_j U_k^\dag U_l^\dag)}
\label{tensor}
\end{align}
and
$\rm Tr$ stands for the summation over $i_n,j_n =1,2,\cdots,K$ 
for all $n \in \Gamma$ 
under the same identification of indices as above.
As shown in Fig.~\ref{fig:tensor},  
the tensor is assigned to each plaquette
and  has four indices corresponding to four links of the plaquette. 

Since our method is based on the discrete approximation with finite $K$, 
we check the convergence of the r.h.s.\ of Eq.~\eqref{TN} 
for large $K$ in actual numerical computations.

In the tensor network \eqref{TN}, 
a single set $\mathring {G}$ is commonly used 
to discretize all the $U_\mu(n)$-integrations. 
Actually, we can use a different set for each link. 
For example, 
tensors can be decomposed in different ways for even and odd sites 
\cite{Levin:2006jai}, 
and we can use four different sets 
$\mathring {G}_1,\mathring {G}_2,\mathring {G}_3,\mathring {G}_4$ 
to discretize the integrations at four links, 
$U_i,\,U_j,\,U_k,\,U_l$, in Fig.~\ref{fig:tensor}. 
We then have 
\begin{align}
  Z \approx {\rm Tr} 
  \prod_{n \in \Gamma_e} T^e_{i_n j_n i'_n j'_n} \cdot 
  \prod_{m \in \Gamma_o} T^o_{i_{m} j_{m} i'_{m} j'_{m}}
\label{TN_Z}
\end{align}
with 
\begin{align}
  T^{e}_{ijkl} \equiv \frac{1}{K^2}\,  
  e^{ -(\beta/N) {\rm Re \, Tr}
  (1-U^{(1)}_i U^{(2)}_j U^{(3)\dag}_k U^{(4)\dag}_l)},
\label{TN_tensor_e}
\\
  T^{o}_{ijkl} \equiv \frac{1}{K^2}\,  
  e^{ -(\beta/N) {\rm Re \, Tr}(1-U^{(3)}_i U^{(4)}_j 
  U^{(1)\dag}_k U^{(2)\dag}_l)},
\label{TN_tensor_o}
\end{align}
where $U^{(a)}_i \in \mathring G_a$ $(a=1,2,3,4)$ 
and $\Gamma_{e/o}$ are the set of even and odd sites, respectively. 
The introduction of four different sets significantly improve 
the precision of the results compared to a single set
as presented in section \ref{sec:numerical_results}. 

\begin{figure}[htbp]
  \centering
  \includegraphics[width=130mm]{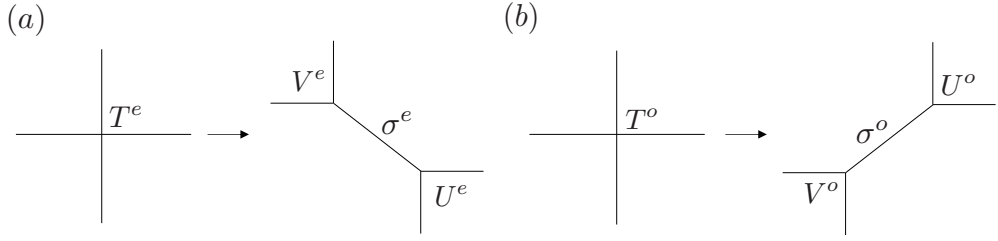}
  \caption{The SVDs of even tensor (a) and odd tensor (b).
}
\label{fig:SVD}
\end{figure}

Once the tensor network is obtained, 
any TRG method can be applied straightforwardly. 
In the Levin-Nave TRG, 
the singular value decomposition (SVD) is employed to decompose the tensors. 
In general, the SVD of an $n \times n$ matrix $M_{ij}$ 
is given by 
\begin{align}
  M_{ij} = \sum_{a=1}^n \sigma_a  U_{ia} V^\ast_{ja},
\end{align}
where $\sigma_a$ are singular values 
sorted as $\sigma_1\ge \sigma_2 \ge \cdots \geq \sigma_n \ge 0$ 
and $U,\,V$ are unitary matrices. 
In our case, 
regarding $T^e_{ijkl}$ (resp.\ $T^o_{ijkl}$) 
as a matrix with the column $ij$ (resp.\ $jk$) 
and the row $kl$ (resp.\ $l i$), 
we have
\begin{align}
  T^e_{ijkl} &= \sum_{A=1}^{K^2} \sigma^e_{A} U^e_{ij,A} V^{e\,\ast}_{kl,A},
\label{svd1}
\\
  T^o_{ijkl} &= \sum_{A=1}^{K^2} \sigma^o_{A} U^o_{jk,A} V^{o\,\ast}_{l i,A}.
\label{svd2}
\end{align}
Figure \ref{fig:SVD} shows these decompositions. 
We again arrive at the tensor network of two-dimensional square lattice 
by defining the renormalized tensor $T^{(1)}$ with bond dimension $D$ as 
\begin{align}
  T^{(1)}_{A_1 A_2 A_3 A_4} =  
  \sqrt{\sigma^e_{A_1} \sigma^o_{A_2} \sigma^e_{A_2} \sigma^o_{A_4} }
  \sum_{i,j,k,l=1}^{D} U^e_{ij,A_1} V^{o\,\ast}_{jk,A_2} V^{e\,\ast}_{kl,A_3} U^o_{li,A_4}.
\label{new_tensor}
\end{align}
The tensor network is repeatedly renormalized in this way. 
Since the bond dimension of the initial tensors 
[Eqs.~\eqref{TN_tensor_e} and \eqref{TN_tensor_o}] is $K$, 
the cost of the first SVD scales with $O(K^6)$. 
Once the tensors are renormalized,  
the bond dimension changes to $D$. 
The cost of the subsequent iterations then scales with $O(D^6)$. 

%%%%%%%%%%%%%%%%%%%%%%%%%%%%%%%%%%%%%%%
\subsection{Relation with the character expansion}
\label{sec:character}
%%%%%%%%%%%%%%%%%%%%%%%%%%%%%%%%%%%%%%%

To understand the group theoretic structure of the SVD 
in the previous subsection, 
we consider the limit $K\to\infty$, 
i.e., the case where the tensor indices continuously take 
all the values  in $G$. 
See appendix \ref{sec:formulas} for a mathematical material 
necessary for the argument below. 

Let $R$ be an irreducible unitary representation of $G$ 
with dimension $d_R$, 
and $D_R(U) = (D^R_{rs}(U))$ $(r,s=1,2,\ldots,d_R$) 
the representation matrix of $U$. 
Denoting the character of $R$ by $\chi_R(U)$, 
the function $e^{-(\beta/N)\, {\rm Re \, tr\,} (1-U) }$ 
can be expanded as
\begin{align}
  e^{-(\beta/N)\, {\rm Re \, tr\,} (1-U) }
  = \sum_R d_R\, \lambda_R (\beta) \chi_R(U). 
\label{character_exp_P}
\end{align}
Here and hereafter, 
$\sum_R$ stands for the summation 
over the irreducible representations $R$. 
The coefficients $\lambda_R(\beta)$ are given by 
\begin{align}
  \lambda_R (\beta) = 
  \frac{1}{d_R}\,\int dU\,e^{-(\beta/N)\, 
  {\rm Re \, tr\,} (1-U)}\,\chi_R(U^{-1}),
\label{def_lambda_R}
\end{align} 
as can be shown by using Eq.~\eqref{CE_formula1}. 

We again consider the infinite dimensional rank-$4$ tensor 
${\mathfrak T}_{g_1 g_2 g_3 g_4}$ [see Eq.~\eqref{inf_tensor}]. 
By using Eq.~\eqref{character_exp_P}, 
this can be written as 
$\sum_R d_R\, \lambda_R\, \chi_R(g_1 g_2 g_3^{-1} g_4^{-1})$
and decomposed in two ways: 
\begin{align}
  {\mathfrak T}_{g_1 g_2 g_3 g_4} 
  &  = \sum_{A=(R,r,s)} 
  {\mathfrak U}^e_{(g_1,g_2), A}\, \lambda_R\, 
  {\mathfrak V}^{e\,\ast}_{(g_3,g_4), A} 
   = \sum_{A=(R,r,s)} 
  {\mathfrak U}^o_{(g_2,g_3), A}\, \lambda_R\, 
  {\mathfrak V}^{o\,\ast}_{(g_4,g_1), A}
\label{inf_svd}
\end{align}
with 
\begin{align}
 {\mathfrak U}^e_{(g_1,g_2), A} =  {\mathfrak V}^e_{(g_2,g_1), A} 
 = {\mathfrak U}^o_{(g_1,g_2^{-1}), A} 
 = {\mathfrak V}^o_{(g_2,g_1^{-1}), A} 
 \equiv  \sqrt{d_R}\, D^R_{rs}(g_1 g_2). 
\end{align}
The Peter-Weyl theorem (see appendix~\ref{sec:formulas}) states that 
the matrix $W_{g,A} \equiv \sqrt{d_R} \,D^R_{rs}(g)$ is unitary. 
Thus, together with the inequality $\lambda_R\geq 0$,%
\footnote{ %-----
  This can be proved by rewriting Eq.~\eqref{def_lambda_R} to the form
  \begin{align}
    e^{\beta}\, d_R\,\lambda_R 
    &= \int dU\,e^{(\beta/(2 N))\,[{\rm tr\,}U+{\rm tr\,}U^{-1}]}\,
    \chi_R(U^{-1})
    = \sum_{m,n=0}^\infty\,\Bigl(\frac{\beta}{2 N}\Bigr)^{m+n}\,
    \frac{C^{(m,n)}_R}{m!\,n!}.
    \nonumber
  \end{align}
  In fact,  
  $C^{(m,n)}_R \equiv
  \int dU\,[{\rm tr\,}U]^m\, [{\rm tr\,}U^{-1}]^n\,\chi_R(U^{-1})
  = \int dU\,[\chi_N(U)]^m\,[\chi_{\bar N}(U)]^n\,\chi_R(U^{-1})  
  $ is the multiplicity of $R$ 
  in the product representation $N^{\otimes m}\otimes{\bar N}^{\otimes n}$, 
  and thus is a nonnegative integer. 
  ($N$ and $\bar N$ are 
  the fundamental and anti-fundamental representations, respectively.)
} %--------------
we find that the decompositions \eqref{inf_svd} are actually SVDs. 
Then, the new tensor ${\mathfrak T}^{(1)}$ 
[Eq.~\eqref{new_tensor} with $D=\infty$] 
is calculated by following Eqs.~\eqref{svd1}--\eqref{new_tensor}, 
and is found to be%
\footnote{ %-----
  We use the symbol $\delta_{R_1 R_2 \ldots R_k} 
  \equiv \delta_{R_1 R_2}\delta_{R_2 R_3}\cdots\delta_{R_{k-1} R_k}$.
} %--------------
\begin{align}
  {\mathfrak T}^{(1)}_{A_1 A_2 A_3 A_4}
  = \frac{\lambda_{R_1}^2}{d_{R_1}}\, \delta_{R_1 R_2 R_3 R_4}\,
  \delta_{s_1 s_2}\,\delta_{r_2 s_3}\,
  \delta_{r_3 r_4}\,\delta_{s_4 r_1}.
\label{inf_new_tensor}
\end{align}
Once this expression is obtained, 
one can perform the TRG iterations (see Appendix~\ref{sec:derivation}) 
to obtain 
\begin{align}
  Z =
  \sum_{R}\, \lambda_R(\beta)^V.
\label{exact_Z}
\end{align}

Recall that 
the TN representation [Eqs.~\eqref{TN_Z}--\eqref{TN_tensor_o}] 
is a discretization of Eqs.~\eqref{inf_Z} and \eqref{inf_tensor}. 
Note that 
the singular values of the tensor ${\mathfrak T}_{g_1 g_2 g_3 g_4}$
have a degeneracy of $d_R^2$ for each $R$ 
because both $r$ and $s$ in $W_{g,A} = \sqrt{d_R} \,D^R_{rs}(g)$ 
take $d_R$ values. 
This means that 
the singular values $\sigma_A$ of our tensor $T_{ijkl}$ 
[Eqs.~\eqref{svd1} and \eqref{svd2}] 
must have this degeneracy approximately. 
We actually find this approximate degeneracy 
in numerical calculations presented in the next section.

%%%%%%%%%%%%%%%%%%%%%%%%%%%%%%%%%%%%%%%
%%%%%%%%%%%%%%%%%%%%%%%%%%%%%%%%%%%%%%%
\section{Numerical results}
\label{sec:numerical_results}
%%%%%%%%%%%%%%%%%%%%%%%%%%%%%%%%%%%%%%%
%%%%%%%%%%%%%%%%%%%%%%%%%%%%%%%%%%%%%%%

In this section, 
we apply our method to the Yang-Mills theory 
with gauge group $G=SU(N)$ $(N=2,3)$ 
on a periodic square lattice. 
We construct the tensor network 
with four different sets $\mathring {G}_a\, (a=1,2,3,4)$ 
of $K$ random link variables 
[see the discussion after Eqs.~\eqref{TN_Z}--\eqref{TN_tensor_o}]. 
We evaluate the free energy density
$f(\beta) \equiv (1/V) \ln Z(\beta)$ 
with the Levin-Nave TRG, 
and the energy density $e(\beta) \equiv  -(\partial /\partial \beta) f(\beta)$ 
by taking numerical derivatives. 
Note that estimates have statistical errors 
in addition to the systematic errors 
coming from the finiteness of $K$ and bond dimension $D$. 
The statistical errors to be given below 
are obtained from five independent trials.

%%%%%%%%%%%%%%%%%%%%%%%%%%%%%%%%%%%%%%% 
\subsection{$SU(2)$}
%%%%%%%%%%%%%%%%%%%%%%%%%%%%%%%%%%%%%%% 

We first make a detailed analysis for $SU(2)$. 
 
\begin{figure}[ht!]
  \centering
  \includegraphics[width=75mm]{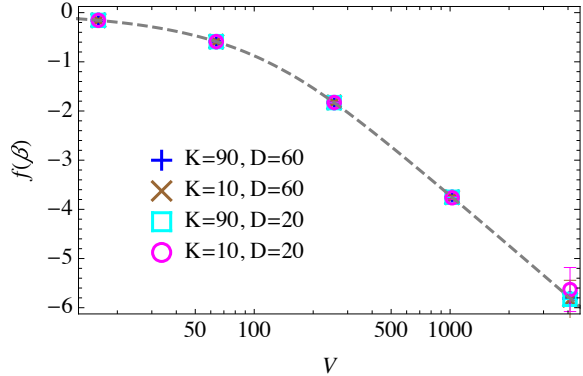}
  \caption{
    \label{fig:f_su2_V}
    Volume dependence of $f(\beta)$ 
    with $\beta/V=0.01$
    for $SU(2)$. 
    The exact values are expressed by the gray dashed line.
  }
\end{figure}

\begin{figure}[ht!]
  \centering
  \includegraphics[width=75mm]{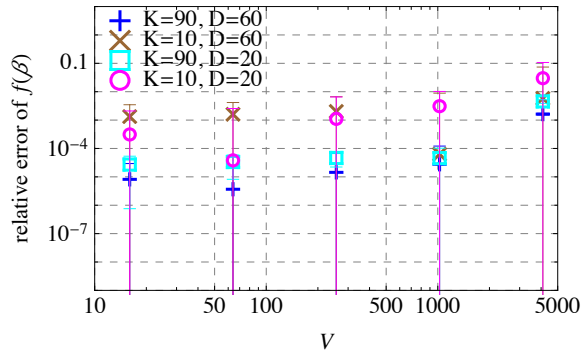}
  \caption{
    \label{fig:df_su2_V}
     Relative error of the free energy density,
     $|f(\beta)-f_{\rm exact}(\beta)|\,/\,|f_{\rm exact}(\beta)|$,
     against volume $V$ 
     with $\beta/V=0.01$ 
     for $SU(2)$. 
  }
\end{figure}

\begin{figure}[ht]
  \centering
  \includegraphics[width=75mm]{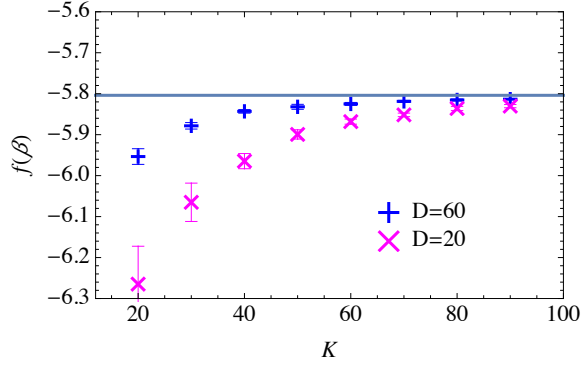}
  \caption{
   \label{fig:f_su2_K}
    $K$ dependence of $f(\beta)$  
    with $\beta/V=0.01$ and $V=64^2$ ($\beta=40.96$) 
    for $SU(2)$. 
  }
\end{figure}

\begin{figure}[ht]
  \centering
  \includegraphics[width=75mm]{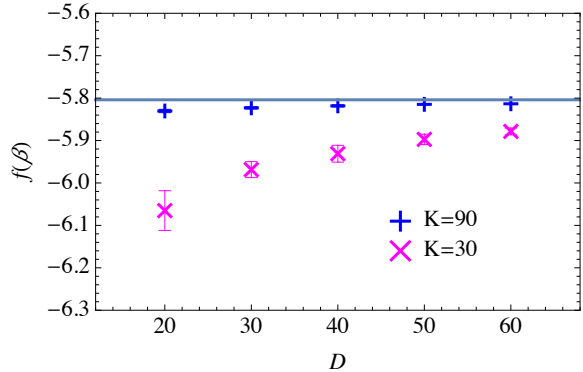}
  \caption{
  \label{fig:f_su2_D}
  $D$ dependence of $f(\beta)$ 
  with $\beta/V=0.01$ and $V=64^2$ ($\beta=40.96$)
  for $SU(2)$. 
  }
\end{figure}

Figure \ref{fig:f_su2_V} shows $f(\beta)$ 
for various volumes $V=L^2$ ($L=4,\,8,\,16,\,32,\,64$) 
with $\beta/V$ fixed to 0.01. 
The exact values are indicated by the gray dashed line. 
Figure~\ref{fig:df_su2_V} shows the relative errors 
to the exact values for the same calculation. 
We see that the numerical results agree well with the exact values. 
We also see that 
as $V$ (and thus $\beta$) is increased, 
larger $K$ and $D$ are required to decrease the systematic errors. 
Figures~\ref{fig:f_su2_K} and \ref{fig:f_su2_D}
show the $K$, $D$ dependences of the free energy density at $V=64^2$ 
($\beta=40.96$). 
We confirm that 
the numerical estimates approach the exact value 
in the limit $K\rightarrow\infty$ and $D\rightarrow\infty$.

\begin{figure}[ht]
  \centering
  \includegraphics[width=75mm]{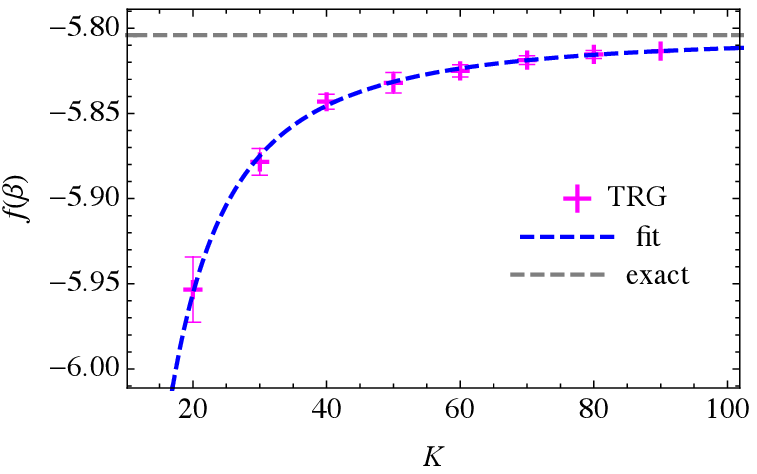} 
  \caption{
    \label{fig:fit_f_su2}
    $\chi^2$ fit of the free energy densities $f(\beta)$ 
    for various $K$ 
    with $\beta/V = 0.01$, $V = 64^2$ and $D=60$ 
    for $SU(2)$. 
  }
\end{figure}

Having obtained the estimates for several values of $K$, 
we can make use of extrapolation to obtain a better estimate. 
Figure~\ref{fig:fit_f_su2} shows the $\chi^2$ fit 
to the obtained data for $D=60$ with the scaling ansatz 
$g(K) \equiv \mu + \alpha K^{-p}$. 
Here, the fitting parameters $\alpha$, $\mu$ and $p$ 
are determined by minimizing the cost function 
\begin{align}
  \chi^2(\mu, \alpha, p) \equiv 
  \sum_{K=20,30,\cdots,90}
  \frac{ 
  [ f(\beta; K) - g(K) ]^2 
  }{ 
  [ \delta f(\beta; K) ]^2 
  }, 
\label{chi_square}
\end{align}
where $f(\beta;K)$ is the obtained value for each $K$, 
and $\delta f(\beta; K)$ the statistical error.
The value of $\mu$ is then used as the final estimate of $f(\beta)$. 

The results of the fitting are summarized in Table \ref{table:su2_fit}. 
\begin{table}[ht]
  \centering
  %-----------------
  \begin{small}
    \begin{tabular}{|c|c|c|c|c|c|}
      \hline
      & (exact)& $\mu$ & $\alpha$& $p$ & $\chi^2/{\rm DOF}$ \\
      \hline
      $f(\beta)$
      & -5.8040 & $-5.8045^{+0.0040}_{-0.0029}$ & $-43^{+26}_{-60}$ & 
        $1.88^{+0.27}_{-0.28}$ &  0.13 \\
      $e(\beta)$
      & 0.03639 & $0.03655^{+0.00029}_{-0.00052}$ & $5^{+24}_{-5}$ & 
        $2.00^{+0.49}_{-0.53}$ &  0.11 \\
      \hline
    \end{tabular}
  \end{small}
  % -----------------
  \caption{Results of the $\chi^2$ fit \eqref{chi_square} for $SU(2)$.}
  \label{table:su2_fit}
\end{table}%
We obtain 
$\mu = -5.8045^{+0.0040}_{-0.0029}$, 
which agrees well with the exact value $f_{\rm exact}(\beta) = -5.8040$. 
Since the estimate without extrapolation is given by  
$f(\beta;K=90) \approx -5.81365 \pm 0.00032$, 
we see that the extrapolation significantly improves the accuracy. 

\begin{figure}[ht]
  \centering
  \includegraphics[width=75mm]{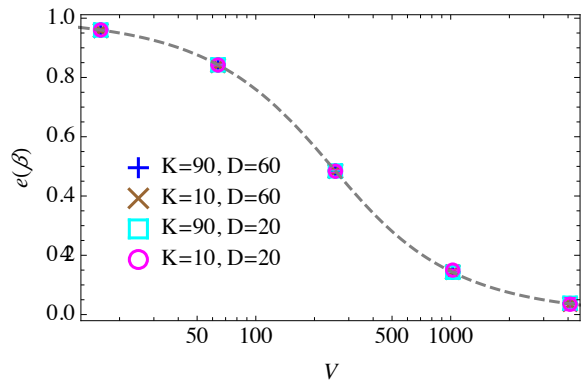}
  \caption{
    \label{fig:e_su2_V}
    Volume dependence of $e(\beta)$ 
    with $\beta/V=0.01$ for $SU(2)$. 
  }
\end{figure}

\begin{figure}[ht]
  \centering
  \includegraphics[width=75mm]{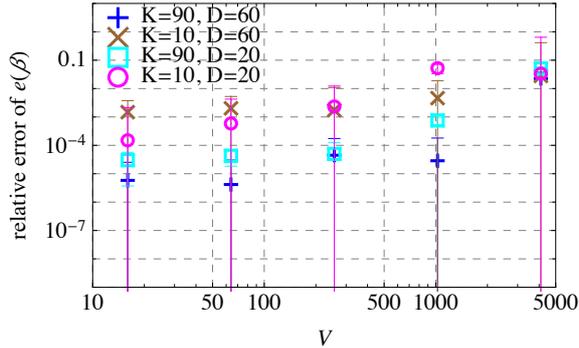}
  \caption{
    \label{fig:de_su2_V}
     Relative error of the energy density,
     $|e(\beta)-e_{\rm exact}(\beta)|\,/\,|e_{\rm exact}(\beta)|$,
     with $\beta/V=0.01$ 
     for $SU(2)$. 
  }
\end{figure}

\begin{figure}[ht]
  \centering
  \includegraphics[width=75mm]{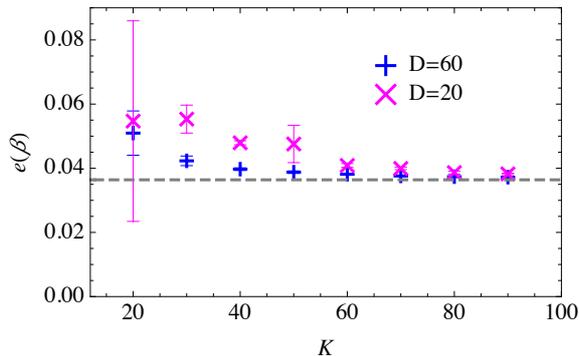}
  \caption{
    \label{fig:e_su2_K}
    $K$ dependence of $e(\beta)$ 
    with $\beta/V=0.01$
    for $SU(2)$. 
  }
\end{figure}

\begin{figure}[ht]
  \centering
  \includegraphics[width=75mm]{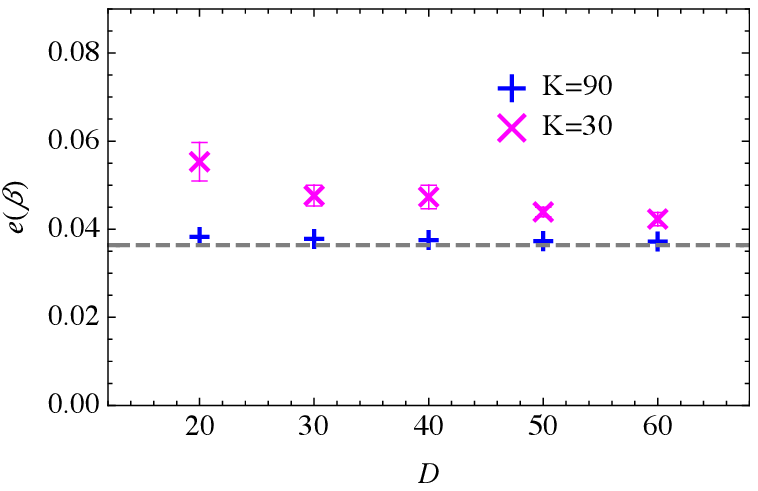}
  \caption{
    \label{fig:e_su2_D}
    $D$ dependence of $e(\beta)$ with $\beta/V=0.01$
    for $SU(2)$. 
  }
\end{figure}

\begin{figure}[ht]
  \centering
  \includegraphics[width=75mm]{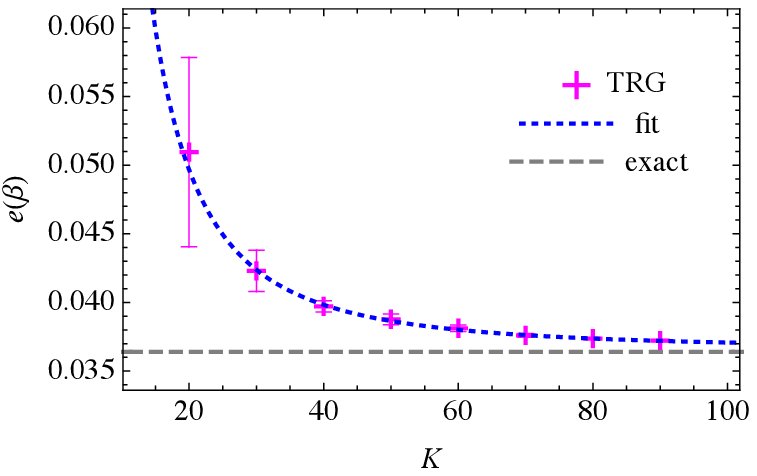} 
  \caption{
    \label{fig:fit_e_su2}
    Fitting of the energy densities $e(\beta)$ 
    for various $K$ 
    with $\beta/V = 0.01$, $V = 64^2$ and $D=60$ 
    for $SU(2)$. 
  }
\end{figure}

We now show the results for the energy density $e(\beta)$. 
In Fig.~\ref{fig:e_su2_V}, 
we plot the estimates of $e(\beta)$ 
for various $V$ with $\beta/V=0.01$ fixed, 
and in Fig.~\ref{fig:de_su2_V} the relative errors to the exact values.
We again see good agreements, 
suggesting the effectiveness of our method. 
In Figs.~\ref{fig:e_su2_K} and \ref{fig:e_su2_D}, 
the $K$ and $D$ dependences are shown for $V=64^2$ $(\beta=40.96)$,  
from which we again confirm that 
the numerical estimates approach the exact value 
in the limit $K\rightarrow\infty$ and $D\rightarrow\infty$.
We can make use of extrapolation to improve the accuracy. 
Figure~\ref{fig:fit_e_su2} shows the $\chi^2$ fit to the obtained data 
with the cost function \eqref{chi_square} 
with $f(\beta)$ replaced by $e(\beta)$. 
The results of the fitting are also given in Table \ref{table:su2_fit}.

\begin{figure}[ht]
  \centering
  \includegraphics[width=86mm]{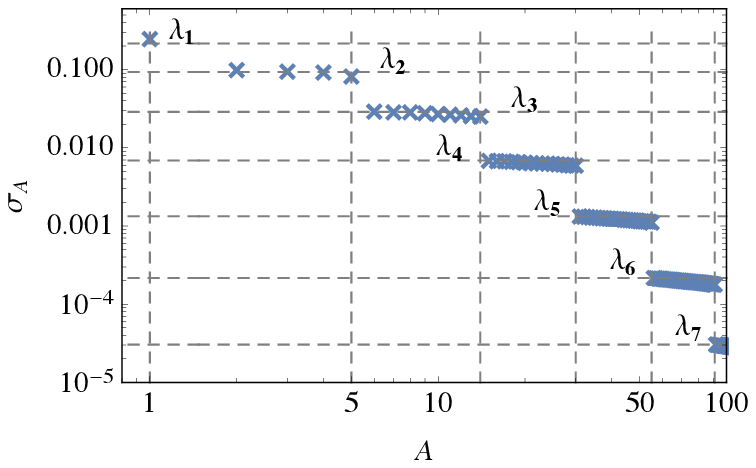}
  \caption{
    \label{fig:sv_su2}
    Singular values $\sigma_A$ of the initial tensors $T^{(e)}$ 
    [Eq.~\eqref{TN_tensor_e}] 
    with $\beta=2$ and $K=90$ 
    for $SU(2)$. 
    Horizontal lines indicate the exact values 
    of $\lambda_R(\beta=2)$, 
    and vertical lines the points 
    at which the exact values change discontinuously. 
  }
\end{figure}

Figure~\ref{fig:sv_su2} shows 
the singular values $\sigma_A$ of the initial tensor 
for $\beta = 2$ with $K=90$.
For $SU(2)$, 
$\lambda_R$ in Eq.~\eqref{character_exp_P} takes the following form 
(see appendix \ref{sec:su(n)}): 
\begin{align}
  \lambda_{R=\boldsymbol{n}}(\beta) = \frac{2}{\beta}\,e^{-\beta} I_n(\beta). 
\end{align}
Here, $\boldsymbol{n}$ is the $n$-dimensional irreducible representation 
of $SU(2)$, 
and $I_n(z)$ is the modified Bessel function of the first kind. 
According to the discussion in Sec.~\ref{sec:character}, 
there will be $d_{R}^2 = n^2$ degenerate singular values 
in the limit $K\rightarrow\infty$ 
for each representation $R=\boldsymbol{n}$. 
In the figure, 
we clearly observe this degeneracy even for finite $K$ ($K=90$ here). 

\begin{figure}[ht]
  \centering
  \includegraphics[width=86mm]{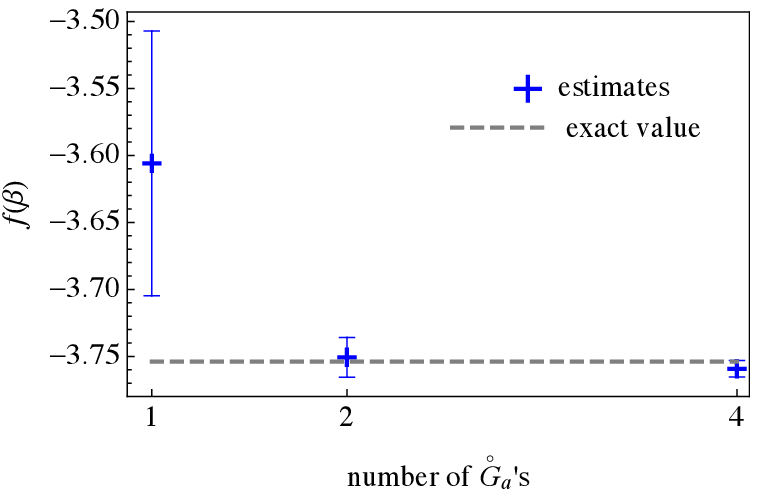}
  \caption{
    \label{fig:Gnum_dependence}
    Dependence of the estimate of $f(\beta)$ 
    on the number of $\mathring{G}_a$'s 
    ($\beta/V=0.04$, $V=16^2$, $K=20$, $D=20$) 
    for $SU(2)$.
    The statistical errors decrease as the number increases. 
  }
\end{figure}

Finally, Fig.~\ref{fig:Gnum_dependence} shows 
the dependence of the estimate of $f(\beta)$ 
on the number of $\mathring{G}_a$'s  
with $\beta/V=0.04$, $V=16^2$, $D=20$, 
from which we see that 
the statistical errors decrease as the number increases. 
This behavior can be understood as follows. 
We first note that 
group elements enter the tensor only in the form of 
the product of two elements, $U_i\,U_j$, 
as can be seen from Eq.~\eqref{inf_svd}. 
We also note that 
a better approximation is achieved 
when the set of $K^2$ elements, $\{U_i\,U_j\}$ $(i,j=1,\ldots,K)$, 
is closer to the uniform distribution on $G$. 
As the number of $\mathring{G}_a$'s increases, 
the set $\{U_i\,U_j\}$
gets more randomly distributed on $G$,  
which leads to a better estimate of observables 
with smaller statistical errors.

%%%%%%%%%%%%%%%%%%%%%%%%%%%%%%%%%%%%%%%
\subsection{$SU(3)$}
\label{sec:su3}
%%%%%%%%%%%%%%%%%%%%%%%%%%%%%%%%%%%%%%%

We make a similar analysis for $SU(3)$ 
with $\beta/V = 0.005$, $V = 64^2$ and $D=90$. 
The irreducible representations $R$ of $SU(3)$ 
are labeled by two nonnegative integers, $R=[q_1,q_2]$ 
(see appendix \ref{sec:su(n)}), 
for which the dimension is given by $d_R = (q_1+1)(q_2+1)(q_1+q_2+2)/2$. 
The coefficients $\lambda_R(\beta)$ are 
given by the formula \eqref{lambda_R_SU(N)}. 
One can show that 
they are ordered as%
\footnote{ %-----
  We write the irreducible representations $R=[q_1,q_2]$ 
  (see appendix \ref{sec:su(n)}) as
  \begin{align}
    &[0,0]=\mathbf{1}, ~~ [1,0]=\mathbf{3}, ~~ [0,1]=\overline{\mathbf{3}}, ~~
     [1,1]=\mathbf{8}, ~~ [2,0]=\mathbf{6}, ~~ [0,2]=\overline{\mathbf{6}}, ~~
  \nonumber
  \\
    &[2,1]=\mathbf{15}, ~~ [1,2]=\overline{\mathbf{15}}, ~~
    [3,0]=\mathbf{10}, ~~ [0,3]=\overline{\mathbf{10}}, ~~
    \ldots.
  \nonumber
  \end{align}
} %-------------- 
\begin{align}
  \lambda_{\mathbf{1}}
  \,>\,
  \lambda_{\mathbf{3}} = \lambda_{\overline{\mathbf{3}}}
  \,>\,
  \lambda_{\mathbf{8}}
  \,>\,
  \lambda_{\mathbf{6}} = \lambda_{\overline{\mathbf{6}}}
  \,>\,
  \lambda_{\mathbf{15}} = \lambda_{\overline{\mathbf{15}}}
  \,>\,
  \lambda_{\mathbf{10}} = \lambda_{\overline{\mathbf{10}}}
  \,>\,
  \cdots. 
\label{ordering_su3}
\end{align}

\begin{figure}[ht]
  \centering
  \includegraphics[width=75mm]{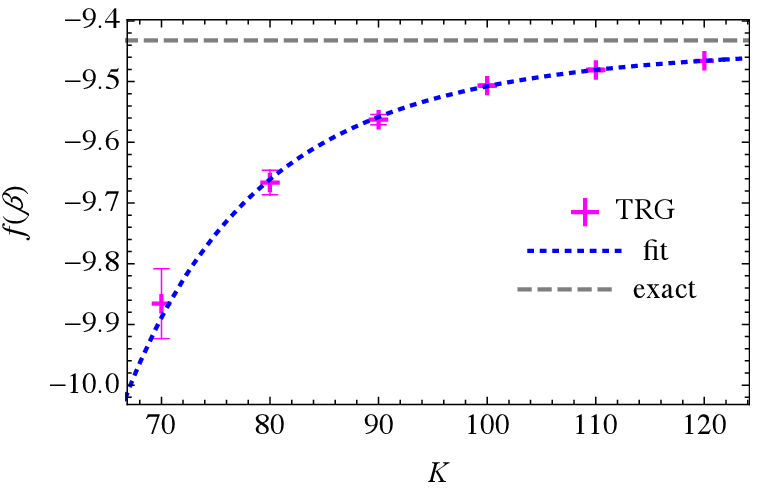} 
  \hspace{5mm}
  \includegraphics[width=75mm]{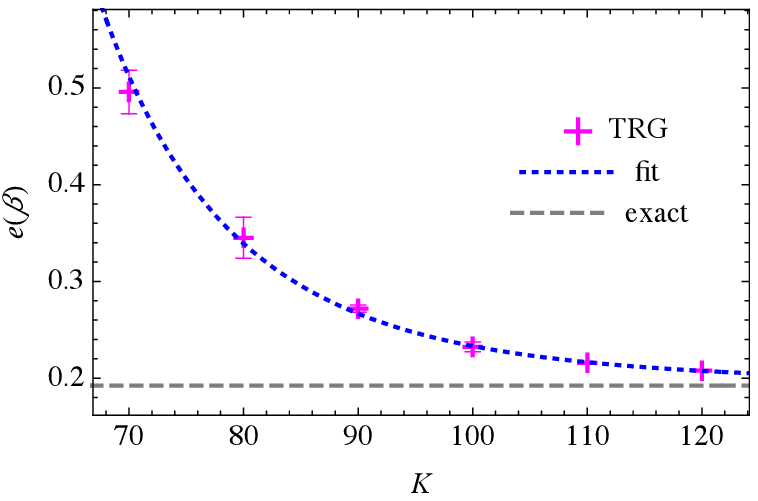} 
  \caption{
    \label{fig:fe_su3}
    $K$ dependences of $f(\beta)$ (left panel) and $e(\beta)$ (right panel) 
    with $\beta/V = 0.005$, $V = 64^2$ and $D=90$ 
    for $SU(3)$. 
  }
\end{figure}

In Fig.~\ref{fig:fe_su3}, 
we plot the free energy densities $f(\beta)$ 
and the energy densities $e(\beta)$ 
against various values of $K$. 
We make the $\chi^2$ fit to the obtained data 
at $K=70,\,80,\ldots,\,120$
again with the scaling ansatz 
$g(K) \equiv \mu + \alpha K^{-p}$. 
A similar analysis is performed for $e(\beta)$. 
The obtained results of the fitting 
are summarized in Table \ref{table:su3_fit}.
\begin{table}[ht]
  \centering
  %-----------------
  \begin{small}
    \begin{tabular}{|c|c|c|c|c|c|}
      \hline
      & (exact)& $\mu$& $\alpha$& $p$& $\chi^2/{\rm DOF}$ \\
      \hline
      $f(\beta)$
      & -9.4323 & $-9.4400^{+0.0019}_{-0.0043}$ & 
        $-0.3^{+0.2}_{-1.7} \times 10^{10}$& $5.31^{+0.44}_{-0.01}$&  0.21 \\
      $e(\beta)$
      & 0.1923 & $0.1941^{+0.0017}_{-0.0008}$ & 
        $2.2^{+5.6}_{-1.6} \times 10^{10}$&  $5.88^{+0.29}_{-0.01}$& 1.18 \\
      \hline
    \end{tabular}
  \end{small}
  % -----------------
  \caption{Results of the $\chi^2$ fit for $SU(3)$.}
  \label{table:su3_fit}
\end{table}%

As for the free energy density $f(\beta)$, 
we obtain the estimate 
$\mu = -9.4400^{+0.0019}_{-0.0043}$, 
which agrees well with the exact value $f_{\rm exact}(\beta)=-9.4323$. 
As for the energy density $e(\beta)$, 
we obtain 
the estimate $\mu = 0.1941^{+0.0017}_{-0.0008}$, 
which also agrees well with the exact value $e_{\rm exact}(\beta)=0.1923$. 
These good agreements show that our method also works for $SU(3)$. 

\begin{figure}[ht]
  \centering
  \includegraphics[width=100mm]{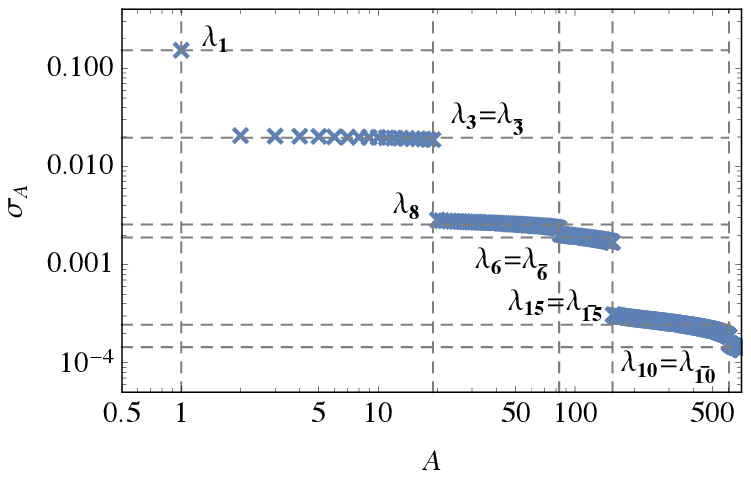}
  \caption{
    \label{fig:sv_su3}
    Singular values $\sigma_A$ of the initial tensors $T^{(e)}$ 
    [Eq.~\eqref{TN_tensor_e}] 
    with $\beta=2$ and $K=120$
    for $SU(3)$. 
    Horizontal lines indicate the exact values 
    of $\lambda_R(\beta=2)$, 
    and vertical lines the points 
    at which the exact values change discontinuously. 
  }
\end{figure}

The singular values of the initial tensor 
agree with the character expansion coefficients 
$\lambda_R(\beta)$ also for $SU(3)$. 
Figure~\ref{fig:sv_su3} shows 
the singular values $\sigma_A$ for $\beta = 2$ with $K=120$. 
We see that 
the coefficients are well reproduced with the correct degeneracies, 
reconfirming the group theoretical structure discussed in Sec.~\ref{sec:character}. 

%%%%%%%%%%%%%%%%%%%%%%%%%%%%%%%%%%%%%%%
%%%%%%%%%%%%%%%%%%%%%%%%%%%%%%%%%%%%%%%
\section{Summary and discussion}
\label{sec:discussion}
%%%%%%%%%%%%%%%%%%%%%%%%%%%%%%%%%%%%%%%
%%%%%%%%%%%%%%%%%%%%%%%%%%%%%%%%%%%%%%%

We have proposed a novel tensor network representation 
for two-dimensional Yang-Mills theories 
with arbitrary compact gauge groups, 
which makes no direct use of the character expansion.
The numerical results for $SU(2)$ and $SU(3)$ gauge groups 
show that our method properly works. 
Although this paper focuses on pure Yang-Mills theories, 
it is straightforward to include the dynamical degrees of freedom 
of fermions and scalar fields into the tensor. 

As a future project, 
it should be important to investigate 
whether the precision is improved 
by applying other renormalization algorithms to our tensor network, 
such as the higher-order tensor renormalization group (HOTRG). 
It should be also interesting to develop a method 
to optimally choose group elements from the group manifold, 
as the Gauss-Hermite quadrature for a field space with flat geometry. 
The extension of the framework to higher-dimensional Yang-Mills theories 
should also be one of the next steps to be considered. 
A study in this direction is now in progress 
and will be reported elsewhere.

%%%%%%%%%%%%%%%%%%%%%%%%%%%%%%%%%%%%%%%
%%%%%%%%%%%%%%%%%%%%%%%%%%%%%%%%%%%%%%%
\section*{Acknowledgments}
This work was partially supported by JSPS KAKENHI 
(Grant Numbers 18J22698, 19K03853, 20H01900) 
and by SPIRITS 
(Supporting Program for Interaction-based Initiative Team Studies) 
of Kyoto University (PI: M.F.). 
D.K.\ would like to thank David C.-J. Lin for encouragement 
and the members of NCTS in National Tsing-Hua University 
for their hospitality. 
N.M.\ is supported by the Special Postdoctoral Researchers Program 
of RIKEN.
%%%%%%%%%%%%%%%%%%%%%%%%%%%%%%%%%%%%%%%
%%%%%%%%%%%%%%%%%%%%%%%%%%%%%%%%%%%%%%%

\appendix

%%%%%%%%%%%%%%%%%%%%%%%%%%%%%%%%%%%%%%%
%%%%%%%%%%%%%%%%%%%%%%%%%%%%%%%%%%%%%%%
\section{Mathematical formulas}
\label{sec:formulas}
%%%%%%%%%%%%%%%%%%%%%%%%%%%%%%%%%%%%%%%
%%%%%%%%%%%%%%%%%%%%%%%%%%%%%%%%%%%%%%%

In this appendix, 
we summarize useful formulas for the integration 
over a compact group $G$. 

For a unitary representation $R$ (not necessarily irreducible) 
with dimension $d_R$, 
we denote the representation matrix of $U\in G$ 
by $D_R(U) = (D^R_{rs}(U))$ $(r,s=1,\ldots,d_R)$ 
and the character 
by $\chi_R(U) = {\rm tr}\,D_R(U)$. 
Note that $\chi_R(1)=d_R$. 
Hereafter we use the term ``representation'' 
as meaning ``representation class'', 
and fix a representative $R$ for each representation class. 
Note that for a unitary representation, 
we have 
$D^R_{rs}(U^{-1})=[D^R_{sr}(U)]^\ast$ 
and $\chi_R(U^{-1})=[\chi_R(U)]^\ast$. 

We introduce the Haar measure $dU$, 
which is two-side invariant and normalized: 
\begin{align}
  \int dU\,f(g_1 U g_2) &= 
  \int dU\,f(U) \quad (\forall g_1,\,g_2\in G),
\\
  \int dU\,f(U^{-1}) &= \int dU\,f(U),
\\
  \int dU\,1 &= 1. 
\end{align}
We also introduce the invariant delta function $\delta(U,V)$ 
associated with the Haar measure: 
\begin{align}
  \int dU\,\delta(U,V)\,f(U) &= f(V),
\\
  \delta(g_1 U g_2,\, g_1 V g_2) &= \delta(U,V) \quad 
  (\forall g_1,\,g_2\in G),
\\
  \delta(U^{-1},V^{-1}) &= \delta(U,V).
\end{align}

We write the set of irreducible unitary representations 
by ${\rm Irrep}=\{R~\mbox{: irreducible}\}$. 
Then, we have the following formula for $R_1,\,R_2\in{\rm Irrep}$:% 
\footnote{ %-----
  From this equation, 
  one can show the formula 
  \begin{align*}
    \int dU\, D^{R_1}_{r_1 s_1} (g_1 U) D^{R_2}_{r_2 s_2} (U^{-1} g_2)
    = \int dU\, D^{R_1}_{r_1 s_1} (g_1 U^{-1}) D^{R_2}_{r_2 s_2} (U g_2)  
    = \frac{\delta_{R_1R_2}}{d_{R_1}}\,
    \delta_{s_1 r_2}\, D^{R_1}_{r_1s_2} (g_1 g_2).
  \end{align*}
} %-------------- 
\begin{align}
  \int dU\, D^{R_1}_{r_1 s_1} (U) D^{R_2}_{r_2 s_2} (U^{-1})  
  =\frac{\delta_{R_1R_2}}{d_{R_1}}\,
  \delta_{r_1 s_2}\, \delta_{s_1 r_2},
\label{master_formula}
\end{align}
from which we readily obtain the formulas 
for the integration of characters,
\begin{align}
  \int dU \, \chi_{R_1} (g_1 U)\, \chi_{R_2}(U^{-1} g_2) 
  &= \frac{\delta_{R_1 R_2}}{d_{R_1}}\, \chi_{R_1}(g_1 g_2),
\label{CE_formula1}
\\
  \int dU \, \chi_{R} (g_1 U g_2\, U^{-1}) 
  &= \frac{1}{d_R}\, \chi_R(g_1)\, \chi_R(g_2).  
\label{CE_formula2}
\end{align}  

The characters of irreducible representations 
$\{\chi_R(U)\}$ $(R\in{\rm Irrep})$ 
form a linear basis of the set of class functions $\{f(U)\}$
that satisfy $f(g U g^{-1}) = f(U)$ $(\forall g\in G)$. 
In particular, 
as can be easily proved, 
$\delta(U,1)$ is expanded as $\sum_{R\in{\rm Irrep}} d_R \chi_R(U)$, 
and thus we have 
\begin{align}
  \delta(U,V) = \sum_{R\in{\rm Irrep}} d_R \chi_R(U V^{-1})
  = \sum_{R\in{\rm Irrep}} d_R \chi_R(V U^{-1}).  
\end{align}
From this equation readily follows the Peter-Weyl theorem, 
which states that the infinite dimensional matrix 
\begin{align}
  W_{U,A} \equiv \sqrt{d_R}\, D^R_{rs}(U)\quad[A=(R,r,s)]
\end{align}
is unitary: 
\begin{align}
  \int dU\, W^\ast_{U,A}\, W_{U,A'}  = \delta_{AA'},\quad
  \sum_A W_{U,A}\, W^\ast_{U',A} = \delta(U,U')
\end{align}
with $\delta_{AA'}\equiv\delta_{RR'}\,\delta_{rr'}\,\delta_{ss'}$ 
and $\sum_{A} \equiv 
\sum_{R\in{\rm Irrep}}\sum_{r=1}^{d_R}\sum_{s=1}^{d_R}$.

%%%%%%%%%%%%%%%%%%%%%%%%%%%%%%%%%%%%%%%
%%%%%%%%%%%%%%%%%%%%%%%%%%%%%%%%%%%%%%%
\section{$\lambda_R(\beta)$ for $G=SU(N)$}
\label{sec:su(n)}
%%%%%%%%%%%%%%%%%%%%%%%%%%%%%%%%%%%%%%%
%%%%%%%%%%%%%%%%%%%%%%%%%%%%%%%%%%%%%%%

\begin{figure}[ht]
  \centering
  \includegraphics[width=60mm]{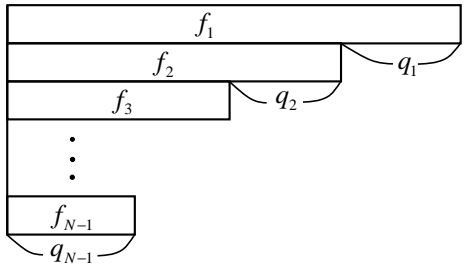} 
  \caption{
    \label{fig:young}
    Young diagram for $R=[q_1,\ldots,q_{N-1}]$.
  }
\end{figure}

For $G=SU(N)$, 
the irreducible representation $R=[q_1,\ldots,q_{N-1}]$ 
($q_i\in\mathbb{Z}_{\geq0}$: Dynkin labels) 
can be labeled by a Young diagram $Y=(f_1,f_2,\ldots,f_{N-1})$ 
($f_1\geq f_2\geq f_{N-1}\geq 0$) 
with the relations $f_i\equiv \sum_{j=i}^{N-1}q_j$ 
(see Fig.~\ref{fig:young}). 
The dimension $d_R$ is given by 
\begin{align}
  d_R = \Delta(\ell_1,\ell_2,\ldots,\ell_{N-1},\ell_N)
  /\Delta(N-1,N-2,\ldots,1,0),
\end{align}
where $\ell_i\equiv f_i+N-i$ with $f_N\equiv 0$ 
and $\Delta(x_1,\ldots,x_N)\equiv \prod_{i<j}(x_i-x_j)$. 
One can show that the coefficients $\lambda_R(\beta)$ 
can be expressed as (see, e.g., \cite{Carlsson:2008dh})
\begin{align}
  \lambda_R(\beta) = \frac{e^{-\beta}}{d_R}\,
  \sum_{Q\in\mathbb{Z}} \det\bigl[
  I_{f_j+i-j+Q}(\beta/N)\bigr]
  \quad [G=SU(N)],
\label{lambda_R_SU(N)}
\end{align}
where $I_n(z)$ are the modified Bessel functions of the first kind. 

For $G=SU(2)$, 
the irreducible representation $R=[q]$ 
corresponds to the spin $j=q/2$ representation 
with $d_R=q+1=2j+1$, 
for which the infinite series \eqref{lambda_R_SU(N)} 
can be summed up to a simple form, 
\begin{align}
  \lambda_R(\beta) = (2/\beta)\,e^{-\beta}\,I_{2j+1}(\beta)
  \quad [G=SU(2)]. 
\label{lambda_R_SU(2)}
\end{align}
Thus, the free energy density and the energy density 
can be expressed as 
\begin{align}
  f(\beta) &={} 
  \frac{1}{V} \log \Big[ \sum_{n = 1} ^\infty 
  \Big(\frac{2}{\beta}\,e^{-\beta}\,I_n(\beta)\Big)^V \Big],\\ 
  e(\beta) &={} -\frac{ \sum_{n = 1} ^\infty
      I_n^{V-1}(\beta)\big[ \big(I_{n+1}(\beta)
      +I_{n-1}(\beta) \big)/2 - I_n(\beta)/\beta \big] }
      {\sum_{n=1}^\infty I_n^V(\beta)}
      + 1 .
\end{align}

%%%%%%%%%%%%%%%%%%%%%%%%%%%%%%%%%%%%%%%
%%%%%%%%%%%%%%%%%%%%%%%%%%%%%%%%%%%%%%%
\section{TN derivation of the exact partition function}
\label{sec:derivation}
%%%%%%%%%%%%%%%%%%%%%%%%%%%%%%%%%%%%%%%
%%%%%%%%%%%%%%%%%%%%%%%%%%%%%%%%%%%%%%%

\begin{figure}[ht]
  \centering
  \includegraphics[width=65mm]{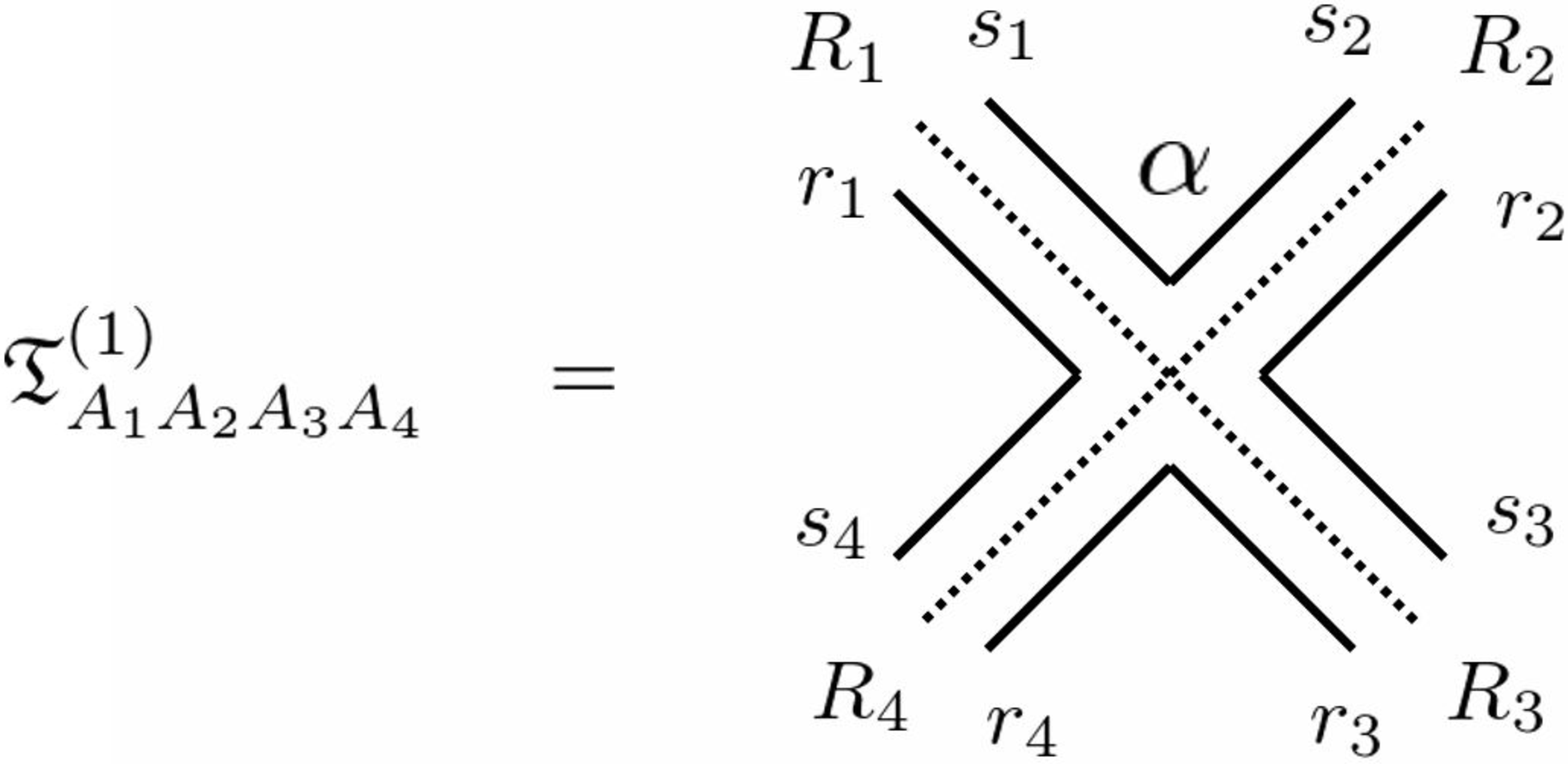}
  \caption{
    \label{fig:T1}
    Graphical representation of  ${\mathfrak T}^{(1)}_{A_1 A_2 A_3 A_4}$. 
  }
\end{figure}

The well-known  formula 
\eqref{exact_Z} can be easily derived  from the TN representation of the partition function
with the infinite dimensional tensor, Eq.~\eqref{inf_new_tensor}: 
\begin{align}
  {\mathfrak T}^{(1)}_{A_1 A_2 A_3 A_4}
  = \alpha_1\, \delta_{R_1 R_2 R_3 R_4}\,
  \delta_{s_1 s_2}\,\delta_{r_2 s_3}\,
  \delta_{r_3 r_4}\,\delta_{s_4 r_1}. 
  \label{mod_inf_T}
\end{align}
Here, $A_i=(R_i,r_i,s_i)$, 
$\delta_{R_1R_2\ldots R_m} = \delta_{R_1R_2} \delta_{R_2R_3}\ldots \delta_{R_{m-1}R_m}$, 
and 
$\alpha_n \equiv \lambda_{R}^{2^n} / d_{R} $ $(n=1,2,\ldots)$ are factors 
located at vertices. 
Figure \ref{fig:T1} shows a graphical representation of ${\mathfrak T}^{(1)}$.

\begin{figure}[htp!]
  \centering
  \includegraphics[width=135mm]{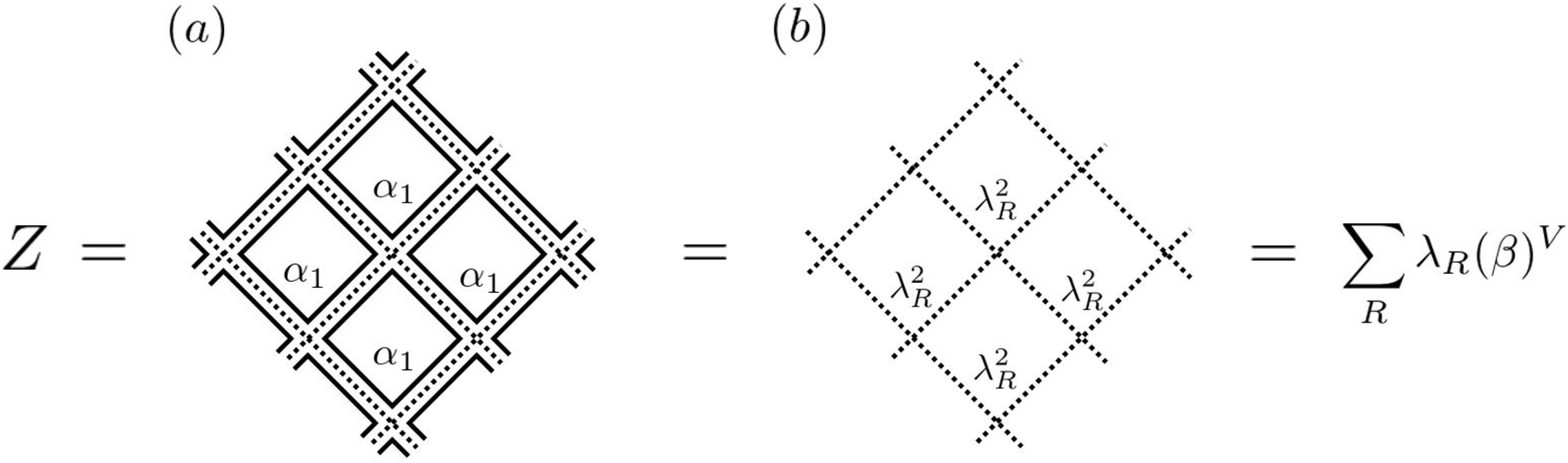}
  \caption{
    \label{fig:Z_inf_TN}
    The TN representation of partition function with the infinite dimensional tensor. 
  }
\end{figure}

It is straightforward to evaluate the value of $Z$ as shown in Fig.~\ref{fig:Z_inf_TN}. 
Figure~\ref{fig:Z_inf_TN} (b) is obtained from Fig.~\ref{fig:Z_inf_TN} (a) 
where $\alpha_1$ is replaced by $\alpha_1 d_R = \lambda_R^2$ 
because $d_R$ is provided from the inner loop. 
The final expression is immediately obtained 
because the remaining tensors in Fig.~\ref{fig:Z_inf_TN} (b) 
are diagonal with respect to the $R$ indices
\cite{Bazavov:2019qih}.

Instead, we can use the TRG iterations to evaluate $Z$.
Omitting the tensor indices, we write 
\begin{align}
  {\mathfrak T}^{(1)}
  =
  \includegraphics[valign=c,width=25mm]{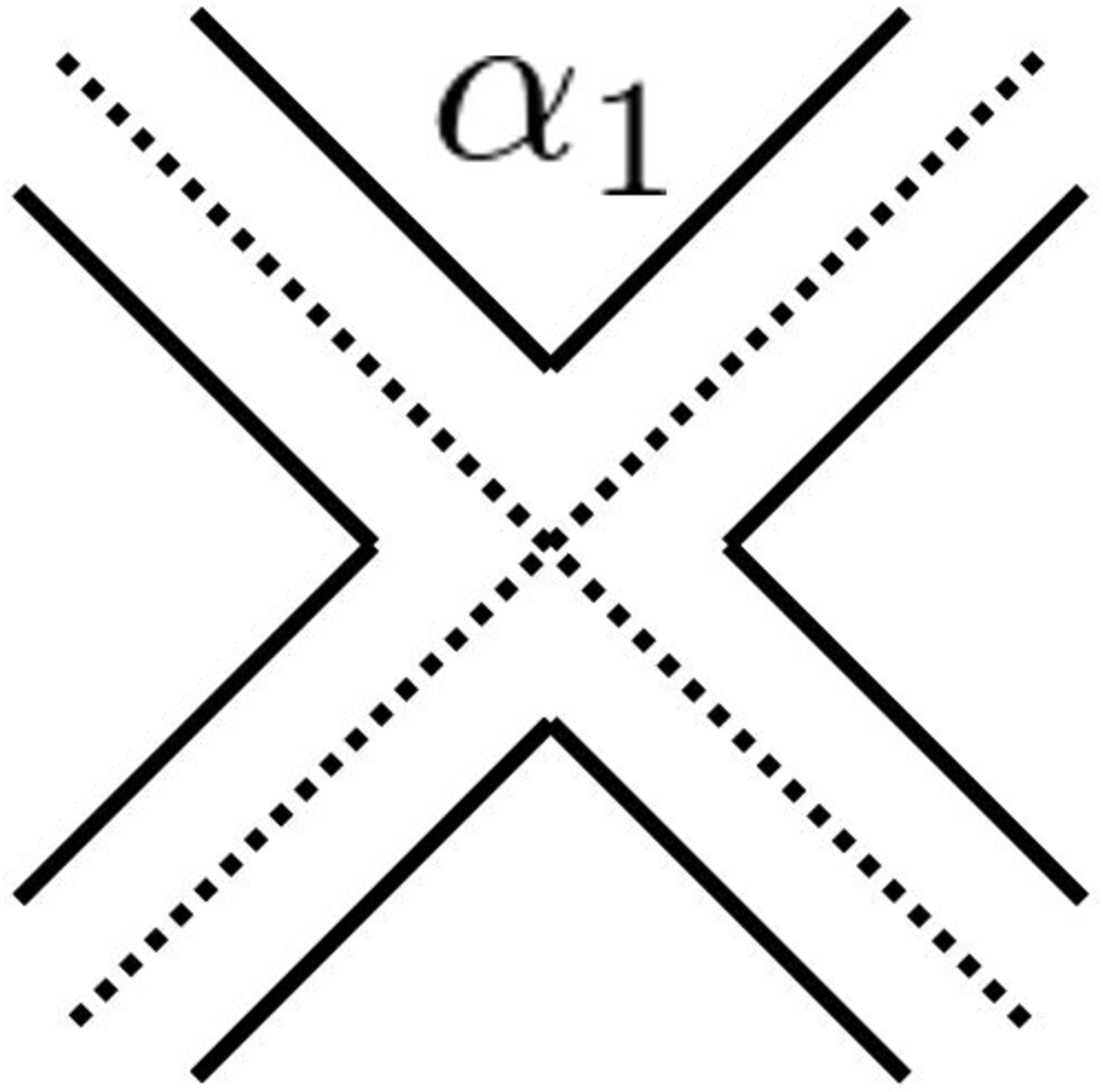}
  ~~.
\label{inf_new_tensor_wo_indices}
\end{align}
Then, we decompose ${\mathfrak T}^{(1)}$ in two ways as 
\begin{align}
  {\mathfrak T}^{(1)}
  = ~~
  \includegraphics[valign=c,width=28mm]{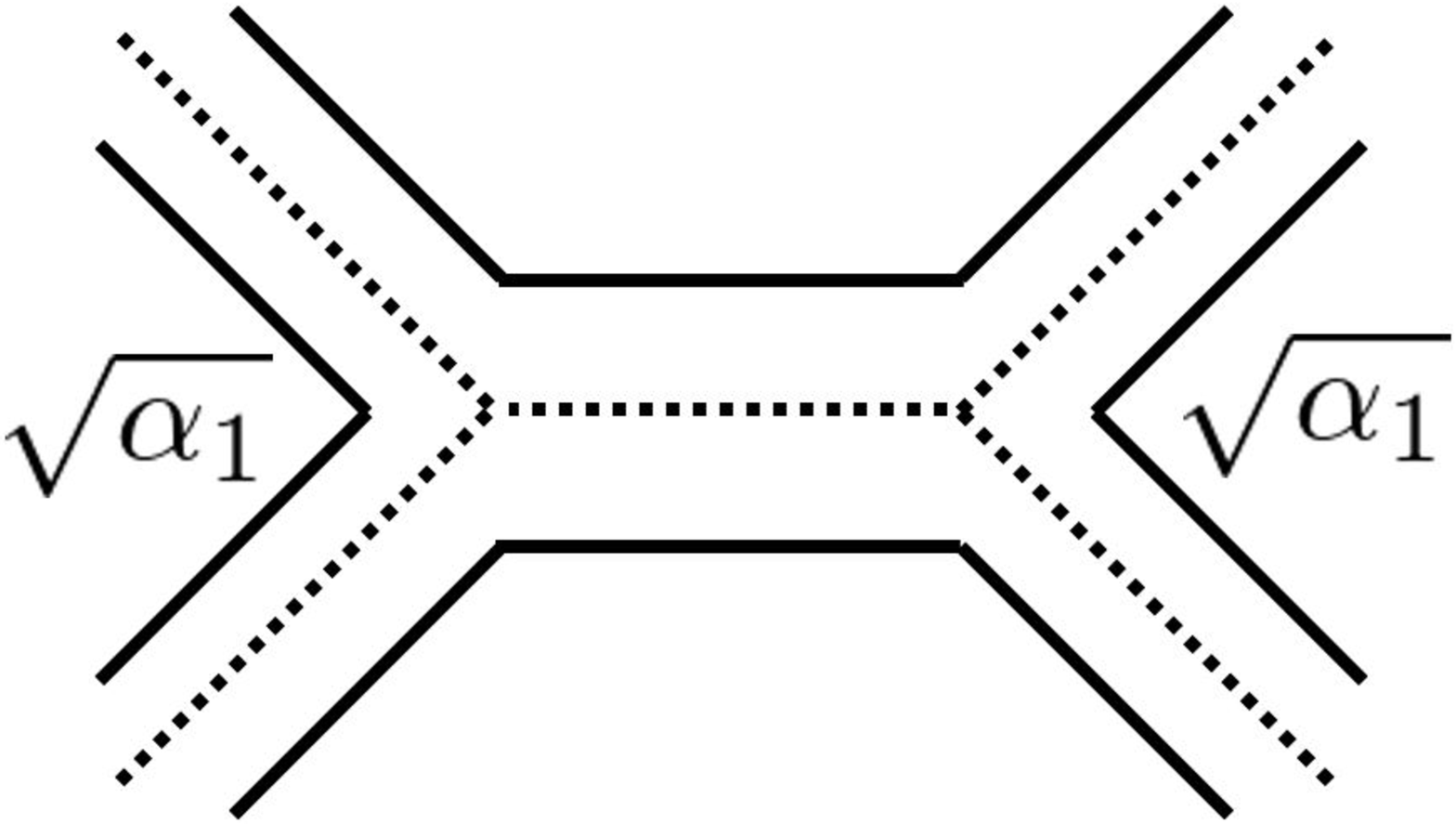}
  \quad = \quad 
  \includegraphics[valign=c,width=19mm]{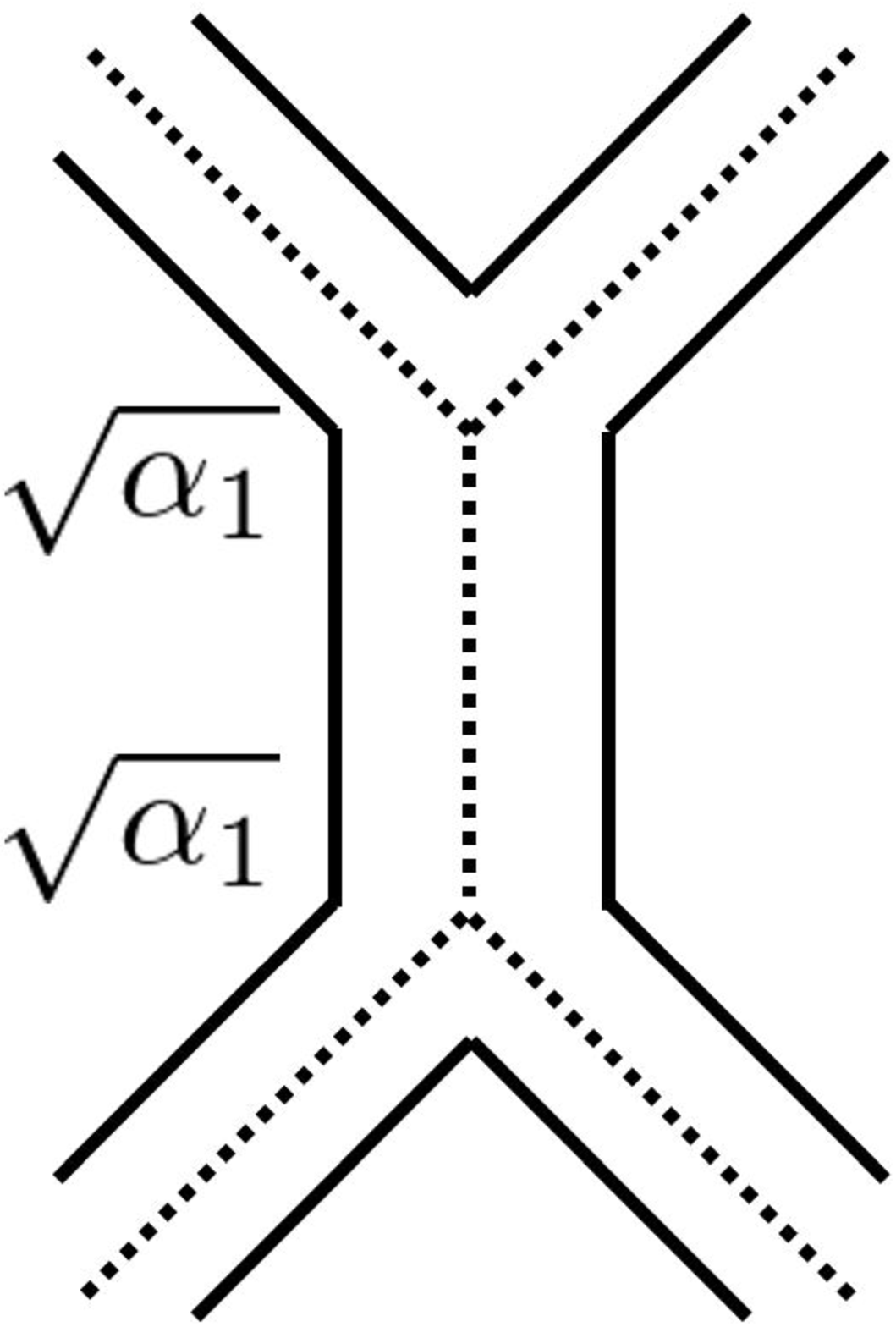}
  ~~,
\label{inf_new_tensor3}
\end{align}
where rank-$3$ tensors are defined in a manner  similar to Eq.~(\ref{mod_inf_T}). 
These decompositions correspond to the SVDs given in Fig.~\ref{fig:SVD}.
With these rank-$3$ tensors, 
we construct the second tensor as 
\begin{align}
  {\mathfrak T}^{(2)} = ~~
  \includegraphics[valign=c,width=25mm]{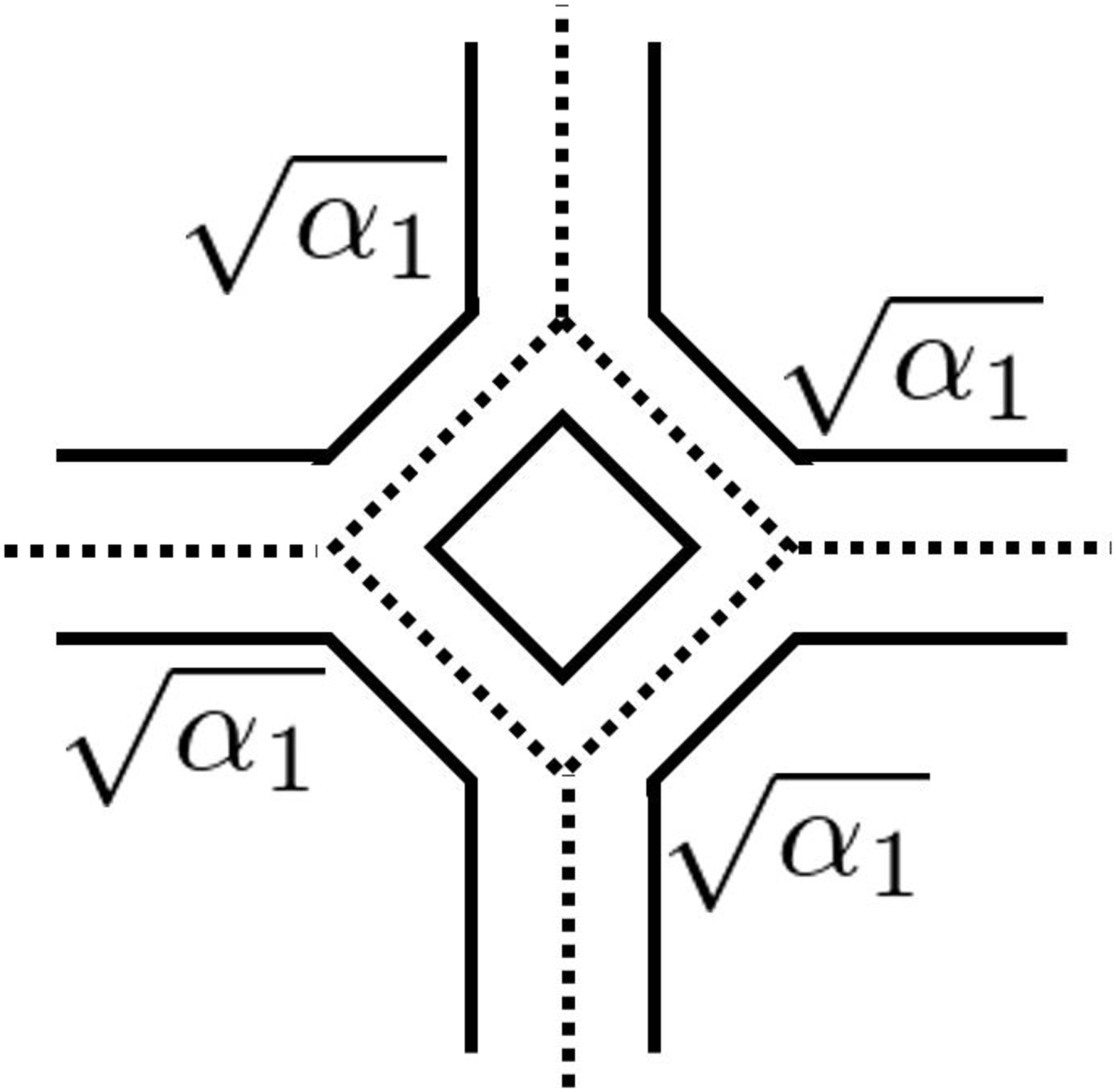}
  ~~=~~
  \includegraphics[valign=c,width=25mm]{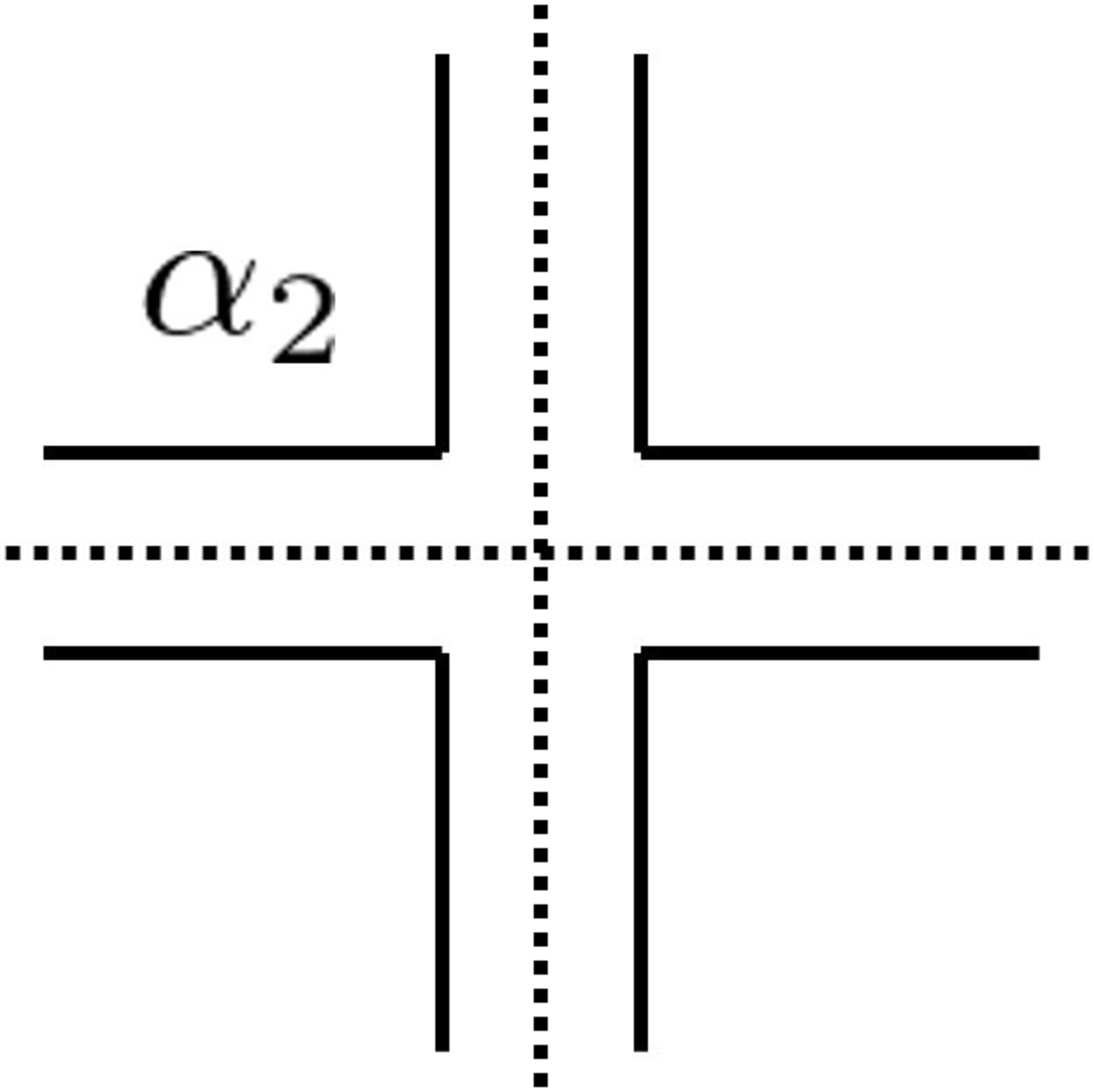} 
  ~~. 
\end{align}
Note that we have not made any truncation. 
Repeating this procedure, 
we have the $n$-th tensor 
\begin{align}
  {\mathfrak T}^{(n)} 
  ~~=~~
  \includegraphics[valign=c,width=25mm]{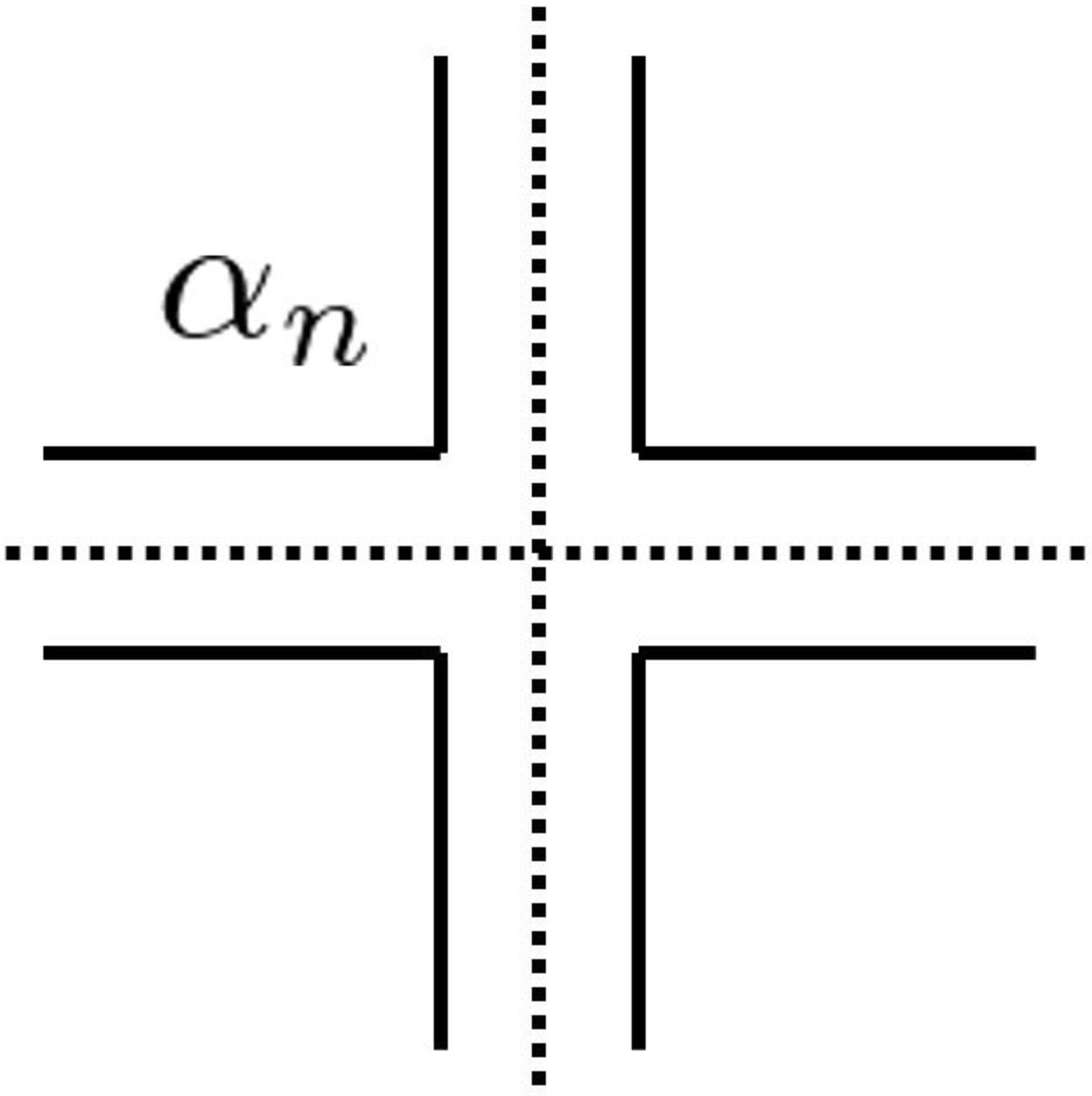} 
  ~~, 
\end{align}
from which the partition function $Z$ 
with volume $V=2^n$ is calculated as 
\begin{align}
  Z =
  \includegraphics[valign=c,width=25mm]{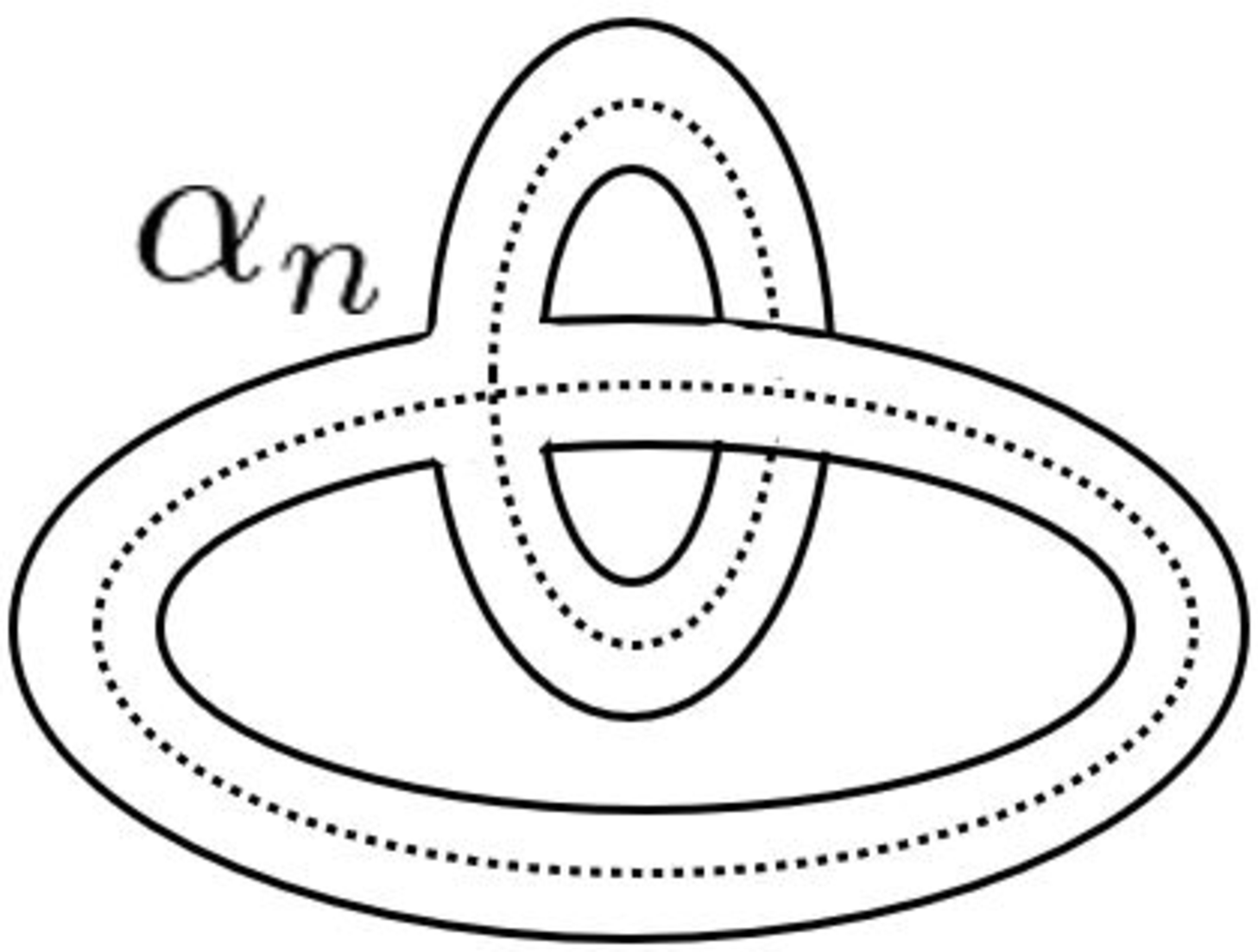} 
  ~~= \sum_{R}\, \lambda_R(\beta)^V. 
\label{inf_exact_Z}
\end{align}

%%%%%%%%%%%%%%%%%%%%%%%%%%%%%%%%%%%%%%%%%%%%%%%%%%%%%%%
\baselineskip=0.9\normalbaselineskip
%%%%%%%%%%%%%%%%%%%%%%%%%%%%%%%%%%%%%%%%%%%%%%%%%%%%%%%

%%%%%%%%%%%%%%%%%%%%%%%%%%%%%%%%%%%%%%%%%%%%%%%%%%%%%%%
%%%%%%%%%%%%%%%%%%%%%%%%%%%%%%%%%%%%%%%%%%%%%%%%%%%%%%%
% \begin{thebibliography}{99}
%   \setlength{\itemsep}{-2pt}
%%%%%%%%%%%%%%%%%%%%%%%%%%%%%%%%%%%%%%%%%%%%%%%%%%%%%%%
%%%%%%%%%%%%%%%%%%%%%%%%%%%%%%%%%%%%%%%%%%%%%%%%%%%%%%%

%%%%%%%%%%%%%%%%%%%%%%%%%%%%%%%%%%%%%%%%%%%%%%%%%%%%%%%
%%%%%%%%%%%%%%%%%%%%%%%%%%%%%%%%%%%%%%%%%%%%%%%%%%%%%%%


\begin{thebibliography}{99}

%\cite{Niggemann:1997cq}
\bibitem{Niggemann:1997cq}
H.~Niggemann, A.~Klumper and J.~Zittartz,
``Quantum phase transition in spin 3/2 systems on the hexagonal lattice: 
Optimum ground state approach,''
Z. Phys. B \textbf{104}, 103-110 (1997)
% doi:10.1007/s002570050425
[arXiv:cond-mat/9702178 [cond-mat]].

%\cite{Verstraete:2004cf}
\bibitem{Verstraete:2004cf}
F.~Verstraete and J.~I.~Cirac,
``Renormalization algorithms for quantum-many body systems 
in two and higher dimensions,''
[arXiv:cond-mat/0407066 [cond-mat]].

%\cite{Levin:2006jai}
\bibitem{Levin:2006jai}
M.~Levin and C.~P.~Nave,
``Tensor renormalization group approach to 2D classical lattice models,''
Phys. Rev. Lett. \textbf{99}, no.12, 120601 (2007)
% doi:10.1103/PhysRevLett.99.120601
[arXiv:cond-mat/0611687 [cond-mat.stat-mech]].

%\cite{Xie:2009zzd}
\bibitem{Xie:2009zzd}
Z.~Y.~Xie, H.~C.~Jiang, Q.~N.~Chen, Z.~Y.~Weng and T.~Xiang,
``Second Renormalization of Tensor-Network States,''
Phys. Rev. Lett. \textbf{103}, 160601 (2009)
% doi:10.1103/PhysRevLett.103.160601
[arXiv:0809.0182 [cond-mat.str-el]].

%\cite{Gu:2010yh}
\bibitem{Gu:2010yh}
Z.~C.~Gu, F.~Verstraete and X.~G.~Wen,
``Grassmann tensor network states and its renormalization 
for strongly correlated fermionic and bosonic states,''
[arXiv:1004.2563 [cond-mat.str-el]].

%\cite{Xie:2012zzz}
\bibitem{Xie:2012zzz}
Z.~Y.~Xie, and J.~Chen, and M.~P.~Qin,and J.~W.~Zhu, and L.~P.~Yang, and T.~Xiang,
``Coarse-graining renormalization by higher-order singular value decomposition,''
Phys. Rev. B \textbf{86}, no.4, 045139 (2012)
% doi:10.1103/physrevb.86.045139
[arXiv:1201.1144 [cond-mat.stat-mech]]. 

%\cite{Adachi:2019paf}
\bibitem{Adachi:2019paf}
D.~Adachi, T.~Okubo and S.~Todo,
``Anisotropic Tensor Renormalization Group,''
Phys. Rev. B \textbf{102}, no.5, 054432 (2020)
% doi:10.1103/PhysRevB.102.054432
[arXiv:1906.02007 [cond-mat.stat-mech]].

%\cite{Kadoh:2019kqk}
\bibitem{Kadoh:2019kqk}
D.~Kadoh and K.~Nakayama,
``Renormalization group on a triad network,''
[arXiv:1912.02414 [hep-lat]].

%\cite{Shimizu:2014uva}
\bibitem{Shimizu:2014uva}
Y.~Shimizu and Y.~Kuramashi,
``Grassmann tensor renormalization group approach 
to one-flavor lattice Schwinger model,''
Phys. Rev. D \textbf{90}, no.1, 014508 (2014)
% doi:10.1103/PhysRevD.90.014508
[arXiv:1403.0642 [hep-lat]].

%\cite{Shimizu:2014fsa}
\bibitem{Shimizu:2014fsa}
Y.~Shimizu and Y.~Kuramashi,
``Critical behavior of the lattice Schwinger model 
with a topological term at $\theta=\pi$ 
using the Grassmann tensor renormalization group,''
Phys. Rev. D \textbf{90}, no.7, 074503 (2014)
% doi:10.1103/PhysRevD.90.074503
[arXiv:1408.0897 [hep-lat]].

%\cite{Shimizu:2017onf}
\bibitem{Shimizu:2017onf}
Y.~Shimizu and Y.~Kuramashi,
``Berezinskii-Kosterlitz-Thouless transition in lattice Schwinger model 
with one flavor of Wilson fermion,''
Phys. Rev. D \textbf{97}, no.3, 034502 (2018)
% doi:10.1103/PhysRevD.97.034502
[arXiv:1712.07808 [hep-lat]].

%\cite{Butt:2019uul}
\bibitem{Butt:2019uul}
N.~Butt, S.~Catterall, Y.~Meurice, R.~Sakai and J.~Unmuth-Yockey,
``Tensor network formulation of the massless Schwinger model 
with staggered fermions,''
Phys. Rev. D \textbf{101}, no.9, 094509 (2020)
% doi:10.1103/PhysRevD.101.094509
[arXiv:1911.01285 [hep-lat]].

%\cite{Takeda:2014vwa}
\bibitem{Takeda:2014vwa}
S.~Takeda and Y.~Yoshimura,
``Grassmann tensor renormalization group 
for the one-flavor lattice Gross\textendash{}Neveu model 
with finite chemical potential,''
PTEP \textbf{2015}, no.4, 043B01 (2015)
% doi:10.1093/ptep/ptv022
[arXiv:1412.7855 [hep-lat]].

%\cite{Akiyama:2020soe}
\bibitem{Akiyama:2020soe}
S.~Akiyama, Y.~Kuramashi, T.~Yamashita and Y.~Yoshimura,
``Restoration of chiral symmetry in cold and dense Nambu--Jona-Lasinio model 
with tensor renormalization group,''
JHEP \textbf{01}, 121 (2021)
% doi:10.1007/JHEP01(2021)121
[arXiv:2009.11583 [hep-lat]].

%\cite{Kadoh:2018tis}
\bibitem{Kadoh:2018tis}
D.~Kadoh, Y.~Kuramashi, Y.~Nakamura, R.~Sakai, S.~Takeda and Y.~Yoshimura,
``Tensor network analysis of critical coupling 
in two dimensional $\phi^{4}$ theory,''
JHEP \textbf{05}, 184 (2019)
% doi:10.1007/JHEP05(2019)184
[arXiv:1811.12376 [hep-lat]].

%\cite{Kadoh:2019ube}
\bibitem{Kadoh:2019ube}
D.~Kadoh, Y.~Kuramashi, Y.~Nakamura, R.~Sakai, S.~Takeda and Y.~Yoshimura,
``Investigation of complex $\phi^{4}$ theory at finite density 
in two dimensions using TRG,''
JHEP \textbf{02}, 161 (2020)
% doi:10.1007/JHEP02(2020)161
[arXiv:1912.13092 [hep-lat]].

%\cite{Akiyama:2020ntf}
\bibitem{Akiyama:2020ntf}
S.~Akiyama, D.~Kadoh, Y.~Kuramashi, T.~Yamashita and Y.~Yoshimura,
``Tensor renormalization group approach 
to four-dimensional complex $\phi^4$ theory at finite density,''
JHEP \textbf{09}, 177 (2020)
% doi:10.1007/JHEP09(2020)177
[arXiv:2005.04645 [hep-lat]].

%\cite{Akiyama:2021zhf}
\bibitem{Akiyama:2021zhf}
S.~Akiyama, Y.~Kuramashi and Y.~Yoshimura,
``Phase transition of four-dimensional lattice $\phi^4$ theory 
with tensor renormalization group,''
[arXiv:2101.06953 [hep-lat]].

%\cite{Asaduzzaman:2019mtx}
\bibitem{Asaduzzaman:2019mtx}
M.~Asaduzzaman, S.~Catterall and J.~Unmuth-Yockey,
``Tensor network formulation of two dimensional gravity,''
Phys. Rev. D \textbf{102}, no.5, 054510 (2020)
% doi:10.1103/PhysRevD.102.054510
[arXiv:1905.13061 [hep-lat]].

%\cite{Bazavov:2019qih}
\bibitem{Bazavov:2019qih}
A.~Bazavov, S.~Catterall, R.~G.~Jha and J.~Unmuth-Yockey,
``Tensor renormalization group study of the non-Abelian Higgs model 
in two dimensions,''
Phys. Rev. D \textbf{99}, no.11, 114507 (2019)
% doi:10.1103/PhysRevD.99.114507
[arXiv:1901.11443 [hep-lat]].

%\cite{Kadoh:2018hqq}
\bibitem{Kadoh:2018hqq}
D.~Kadoh, Y.~Kuramashi, Y.~Nakamura, R.~Sakai, S.~Takeda and Y.~Yoshimura,
``Tensor network formulation for two-dimensional lattice 
$ \mathcal{N} $ = 1 Wess-Zumino model,''
JHEP \textbf{03}, 141 (2018)
% doi:10.1007/JHEP03(2018)141
[arXiv:1801.04183 [hep-lat]].

%\cite{Kawauchi:2016xng}
\bibitem{Kawauchi:2016xng}
H.~Kawauchi and S.~Takeda,
``Tensor renormalization group analysis of CP(N-1) model,''
Phys. Rev. D \textbf{93}, no.11, 114503 (2016)
% doi:10.1103/PhysRevD.93.114503
[arXiv:1603.09455 [hep-lat]].

%\cite{Akiyama:2020sfo}
\bibitem{Akiyama:2020sfo}
S.~Akiyama and D.~Kadoh,
``More about the Grassmann tensor renormalization group,''
[arXiv:2005.07570 [hep-lat]].

%\cite{Kadoh:2021fri}
\bibitem{Kadoh:2021fri}
D.~Kadoh, H.~Oba and S.~Takeda,
``Triad second renormalization group,''
[arXiv:2107.08769 [cond-mat.str-el]].

%\cite{Kuramashi:2019cgs}
\bibitem{Kuramashi:2019cgs}
Y.~Kuramashi and Y.~Yoshimura,
``Tensor renormalization group study 
of two-dimensional U(1) lattice gauge theory with a $\theta$ term,''
JHEP \textbf{04}, 089 (2020)
% doi:10.1007/JHEP04(2020)089
[arXiv:1911.06480 [hep-lat]].

% \cite{Parisi:1983cs}
\bibitem{Parisi:1983cs}
G.~Parisi,
``On complex probabilities,''
Phys.\ Lett.\ B {\bf 131}, 393 (1983).

% \cite{Klauder:1983sp}
\bibitem{Klauder:1983sp}
J.R.~Klauder,
``Coherent State Langevin Equations for Canonical Quantum Systems 
With Applications to the Quantized Hall Effect,''
Phys.\ Rev.\ A {\bf 29}, 2036 (1984).

%\cite{Aarts:2009dg}
\bibitem{Aarts:2009dg}
G.~Aarts, F.~A.~James, E.~Seiler and I.~O.~Stamatescu,
``Adaptive stepsize and instabilities in complex Langevin dynamics,''
Phys.\ Lett.\ B \textbf{687}, 154-159 (2010)
% doi:10.1016/j.physletb.2010.03.012
[arXiv:0912.0617 [hep-lat]].

%\cite{Nishimura:2015pba}
\bibitem{Nishimura:2015pba}
J.~Nishimura and S.~Shimasaki,
``New Insights into the Problem with a Singular Drift Term 
in the Complex Langevin Method,''
Phys.\ Rev.\ D \textbf{92}, no.1, 011501 (2015)
% doi:10.1103/PhysRevD.92.011501
[arXiv:1504.08359 [hep-lat]].

%\cite{Witten:2010cx}
\bibitem{Witten:2010cx}
E.~Witten,
``Analytic continuation of Chern-Simons theory,''
AMS/IP Stud.\ Adv.\ Math.\ \textbf{50}, 347-446 (2011)
[arXiv:1001.2933 [hep-th]].

%\cite{Cristoforetti:2012su}
\bibitem{Cristoforetti:2012su} 
M.~Cristoforetti, F.~Di Renzo and L.~Scorzato,
``New approach to the sign problem in quantum field theories: 
High density QCD on a Lefschetz thimble,''
Phys.\ Rev.\ D {\bf 86}, 074506 (2012)
% doi:10.1103/PhysRevD.86.074506
[arXiv:1205.3996 [hep-lat]].

%\cite{Cristoforetti:2013wha}
\bibitem{Cristoforetti:2013wha} 
M.~Cristoforetti, F.~Di Renzo, A.~Mukherjee and L.~Scorzato,
``Monte Carlo simulations on the Lefschetz thimble: 
Taming the sign problem,''
Phys.\ Rev.\ D {\bf 88}, no.\ 5, 051501(R) (2013)
% doi:10.1103/PhysRevD.88.051501
[arXiv:1303.7204 [hep-lat]].
 
%\cite{Fujii:2013sra}
\bibitem{Fujii:2013sra} 
H.~Fujii, D.~Honda, M.~Kato, Y.~Kikukawa, S.~Komatsu and T.~Sano,
``Hybrid Monte Carlo on Lefschetz thimbles 
- A study of the residual sign problem,''
JHEP {\bf 1310}, 147 (2013)
% doi:10.1007/JHEP10(2013)147
[arXiv:1309.4371 [hep-lat]].

%\cite{Alexandru:2015sua} 
\bibitem{Alexandru:2015sua} 
A.~Alexandru, G.~Ba\c sar, P.~F.~Bedaque, G.~W.~Ridgway and N.~C.~Warrington,
``Sign problem and Monte Carlo calculations beyond Lefschetz thimbles,''
JHEP {\bf 1605}, 053 (2016)
% doi:10.1007/JHEP05(2016)053
[arXiv:1512.08764 [hep-lat]].

%\cite{Fukuma:2017fjq}
\bibitem{Fukuma:2017fjq} 
M.~Fukuma and N.~Umeda,
``Parallel tempering algorithm for integration over Lefschetz thimbles,''
PTEP {\bf 2017}, no.\ 7, 073B01 (2017)
% doi:10.1093/ptep/ptx081
[arXiv:1703.00861 [hep-lat]].

%\cite{Alexandru:2017oyw}
\bibitem{Alexandru:2017oyw} 
A.~Alexandru, G.~Ba\c sar, P.~F.~Bedaque and N.~C.~Warrington,
``Tempered transitions between thimbles,''
Phys.\ Rev.\ D {\bf 96}, no.\ 3, 034513 (2017)
% doi:10.1103/PhysRevD.96.034513
[arXiv:1703.02414 [hep-lat]].

%\cite{Fukuma:2019wbv}
\bibitem{Fukuma:2019wbv} 
M.~Fukuma, N.~Matsumoto and N.~Umeda,
``Applying the tempered Lefschetz thimble method 
to the Hubbard model away from half filling,''
Phys.\ Rev.\ D {\bf 100}, no.\ 11, 114510 (2019)
% doi:10.1103/PhysRevD.100.114510
[arXiv:1906.04243 [cond-mat.str-el]].

%\cite{Fukuma:2019uot}
\bibitem{Fukuma:2019uot}
M.~Fukuma, N.~Matsumoto and N.~Umeda,
``Implementation of the HMC algorithm 
on the tempered Lefschetz thimble method,''
[arXiv:1912.13303 [hep-lat]].
  
% \cite{Fukuma:2020fez}
\bibitem{Fukuma:2020fez}
  M.~Fukuma and N.~Matsumoto,
  ``Worldvolume approach to the tempered Lefschetz thimble method,''
  PTEP \textbf{2021}, no.2, 023B08 (2021)
  % doi:10.1093/ptep/ptab010
  [arXiv:2012.08468 [hep-lat]].

%\cite{Fukuma:2021aoo}
\bibitem{Fukuma:2021aoo}
M.~Fukuma, N.~Matsumoto and Y.~Namekawa,
``Statistical analysis method for the worldvolume hybrid Monte Carlo algorithm,''
[arXiv:2107.06858 [hep-lat]].

%\cite{Mori:2017pne}
\bibitem{Mori:2017pne}
Y.~Mori, K.~Kashiwa and A.~Ohnishi,
``Toward solving the sign problem with path optimization method,''
Phys.\ Rev.\ D \textbf{96}, no.11, 111501 (2017)
% doi:10.1103/PhysRevD.96.111501
[arXiv:1705.05605 [hep-lat]].

%\cite{Mori:2017nwj}
\bibitem{Mori:2017nwj}
Y.~Mori, K.~Kashiwa and A.~Ohnishi,
``Application of a neural network to the sign problem 
via the path optimization method,''
PTEP \textbf{2018}, no.2, 023B04 (2018)
% doi:10.1093/ptep/ptx191
[arXiv:1709.03208 [hep-lat]].

%\cite{Alexandru:2018fqp}
\bibitem{Alexandru:2018fqp}
A.~Alexandru, P.~F.~Bedaque, H.~Lamm and S.~Lawrence,
``Finite-Density Monte Carlo Calculations on Sign-Optimized Manifolds,''
Phys.\ Rev.\ D \textbf{97}, no.9, 094510 (2018)
% doi:10.1103/PhysRevD.97.094510
[arXiv:1804.00697 [hep-lat]].

%\cite{Carlsson:2008dh}
\bibitem{Carlsson:2008dh}
J.~Carlsson,
``Integrals over SU(N),''
[arXiv:0802.3409 [hep-lat]].

\end{thebibliography}
\end{document}